\documentclass[aps,prd,amsmath,amssymb,twocolumn,superscriptaddress,preprintnumbers,nofootinbib]{revtex4-1}

\pdfoutput=1
\usepackage[utf8]{inputenc} 
\usepackage{graphicx}
\usepackage{multirow}
\usepackage{url}
\usepackage[bookmarks, pagebackref=false]{hyperref}
\usepackage[usenames,dvipsnames]{xcolor}
\definecolor{rossoCP3}{cmyk}{0,.88,.77,.40}
\definecolor{blaa}{rgb}{0.2,0.2,0.6}
\hypersetup{colorlinks, 
	bookmarksopen, 
	bookmarksnumbered,
	citecolor=blaa,
	linkcolor=rossoCP3,
	urlcolor=rossoCP3, 
}
\usepackage{bm}
\usepackage{bbm}

\usepackage{amsmath,amssymb,amsfonts,amsthm}
\usepackage{float}
\usepackage[Symbolsmallscale]{upgreek}
\usepackage{dsfont}
\usepackage[vcentermath]{youngtab}

\usepackage{placeins}
\usepackage{xspace}
\usepackage{cancel} 
\usepackage{slashed}
\usepackage{natbib}
\usepackage{booktabs,multirow}

\allowdisplaybreaks

\graphicspath{{./figs/}}

\begin{document}


\title{\Large\color{rossoCP3} Towards the QED beta function and renormalons at \texorpdfstring{$1/N_f^2$}{1Nf2} and \texorpdfstring{$1/N_f^3$}{1Nf3}}

\author{Nicola Andrea {\sc Dondi}}
\thanks{{\scriptsize Email}: \href{mailto:dondi@cp3.sdu.dk}{dondi@cp3.sdu.dk}; {\scriptsize ORCID}: \href{https://orcid.org/0000-0002-6971-2028}{ 0000-0002-6971-2028}}
\affiliation{CP$^3$-Origins,  University of Southern Denmark, Campusvej 55, 5230 Odense M, Denmark}
\affiliation{Institute for Particle Physics Phenomenology, Durham University, South Road, Durham, DH1 3LE}

\author{Gerald V.~{\sc Dunne}}
\thanks{{\scriptsize Email}: \href{mailto:gerald.dunne@uconn.edu}{gerald.dunne@uconn.edu}; {\scriptsize ORCID}: \href{https://orcid.org/0000-0003-1338-339X}{ 0000-0003-1338-339X}}
\affiliation{Physics Department, University of Connecticut, Storrs CT 06269-3046, USA}

\author{Manuel {\sc Reichert}}
\thanks{{\scriptsize Email}: \href{mailto:reichert@cp3.sdu.dk}{reichert@cp3.sdu.dk}; {\scriptsize ORCID}: \href{https://orcid.org/0000-0003-0736-5726}{ 0000-0003-0736-5726}}
\affiliation{CP$^3$-Origins,  University of Southern Denmark, Campusvej 55, 5230 Odense M, Denmark}

\author{Francesco {\sc Sannino}}
\thanks{{\scriptsize Email}: \href{mailto:sannino@cp3.sdu.dk}{sannino@cp3.sdu.dk}; {\scriptsize ORCID}: \href{https://orcid.org/0000-0003-2361-5326}{ 0000-0003-2361-5326}}
\affiliation{CP$^3$-Origins,  University of Southern Denmark, Campusvej 55, 5230 Odense M, Denmark} 
\affiliation{Dipartimento di Fisica “E. Pancini”, Università di Napoli Federico II | INFN sezione di Napoli, Complesso Universitario di Monte S. Angelo Edificio 6, via Cintia, 80126 Napoli, Italy}

\begin{abstract} 
We determine the $1/N_f^2$ and $1/N_f^3$  contributions to the QED beta function stemming from the closed set of nested diagrams.
At order $1/N_f^2$  we discover a new logarithmic branch-cut  closer to the origin when compared to the $1/N_f$ results.
The same singularity location appears at $1/N_f^3$, and these correspond to a UV renormalon singularity in the finite part of the photon two-point function.
\end{abstract}

\maketitle

\section{Introduction}
The discovery of asymptotically safe quantum field theories in four dimensions \cite{Litim:2014uca,Litim:2015iea} triggered renewed interest in studying the ultraviolet fate of quantum field theories once asymptotic freedom is lost. The original proof of asymptotic safety made use of the Veneziano-Witten large number of flavors and colors limit for a class of gauge-Yukawa theories that displayed perturbatively trustable ultraviolet fixed points. Without scalars, it is impossible to analytically disentangle the ultraviolet fate of asymptotically non-free gauge-fermion theories. Nevertheless one can make progress by analyzing the large $N_f$ dynamics of these theories at finite number of colors  \cite{PalanquesMestre:1983zy,Gracey:1996he,Ferreira:1997bi,Holdom:2010qs,Pica:2010xq,Shrock:2013cca,Gracey:2018ame} including again certain type of Yukawa interactions \cite{Mann:2017wzh,Pelaggi:2017abg,Kowalska:2017pkt,Antipin:2018zdg}. These studies make use of the large $N_f$ resummation techniques to derive the all orders in the 't Hooft coupling beta functions of these theories at order $1/N_f$. This large $N_f$ beta function has several interesting properties including the emergence of singularities undermining the consistency of the expansion, whose physical interpretation remains still to be clarified \cite{Alanne:2019vuk}. In the meantime first principle lattice simulations have begun to explore the large $N_f$ dynamics in a systematic manner \cite{Leino:2019qwk}. It is therefore highly desirable to gain insight into the sub-leading $1/N_f^2$ corrections. This task, however, turns out to be challenging. The present work constitutes a step forward in this direction by determining these sub-leading corrections for a closed class of diagrams in QED.

To achieve our goal we will make use of the technologies developed in our recent work \cite{Dondi:2019ivp}  according to which it is shown that it is possible to reconstruct the $1/N_f$ beta function and its properties using a finite number of coefficients of the perturbative series. We determined the stability of the series and showed that about thirty terms were needed to properly reconstruct the $1/N_f$ beta function up to the leading singularity.  The technology includes Pad\'e methods, combined with the study of the large order growth of the perturbative series. 

In this paper, we employ this technology to deduce the first complete set of $1/N_f^2$  and $1/N_f^3$ corrections for the QED nested diagrams. We discover the emergence of a novel singularity  at order $1/N_f^2$ within this subset of diagrams. The latter appears closer to the origin when compared to the original $1/N_f$ singularity of the full beta function. We refrain from speculating about the physical content of this singularity given that the remaining diagrams are still to be computed. The nature of the singularity is captured by a novel logarithmic branch-cut at a value of the 't Hooft coupling where the beta function remains finite. {We also find an intriguing correspondence between the UV renormalons in the finite part of the photon two-point function, appearing at multiples of 3 in the Borel plane, and the leading singularities of the divergent part  of the photon two-point function, at order $1/N_f^2$ and $1/N_f^3$.}

The paper is organized as follows. In \autoref{sec:LNFQED} we fix the notation and introduce the basic building blocks for our computation. This is followed by \autoref{sec:CND}, in which we present details of the computation and uncover the beta function contribution and its leading singularity of the nested diagrams.  In \autoref{sec:renormalons} we study the appearance of renormalons in the finite part of the photon two-point function. We offer our conclusions in \autoref{sec:conclusions}. The details of the various computations can be found in the appendices. 

\section{Large \texorpdfstring{$N_f$}{Nf} QED setup} 
\label{sec:LNFQED}
In QED with a large number of flavors $N_f$, it is natural to introduce the 't Hooft coupling  
\begin{align}
K = \frac{g^2 N_f }{4\pi^2} \,,
\end{align}
which we keep fixed when sending $N_f$ to infinity.
This allows organizing the beta function as a series in $1/N_f$
\begin{align}
   \label{eq:beta}
  \beta(K)  & \equiv \mu \frac{\mathrm d K}{\mathrm d \mu} = \sum_{k=0}^\infty \frac{ \beta^{(k)}(K)}{N_f^k} \,,
\end{align}
where we have introduced the renormalization group (RG) scale $\mu$. The beta function describes the change of the coupling strength with respect to this RG scale.
In \eqref{eq:beta}, each $\beta^{(k)}(K)$ constitutes itself a perturbative expansion in the 't Hooft coupling $K$.
With this counting, a fermion bubble is of order one and each photon line in a diagram is dressed with $n$ fermion bubbles as depicted in the following diagram
\begin{align}
 \parbox{3.62cm}{\includegraphics[scale=.3]{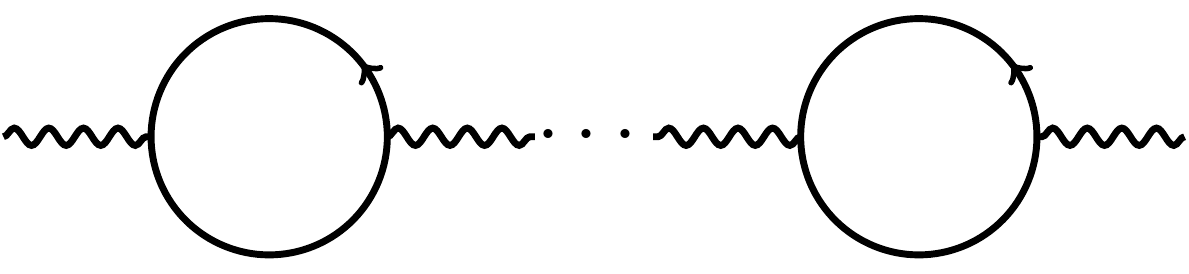}}
  \equiv \parbox{1.8cm}{\includegraphics[scale=.3]{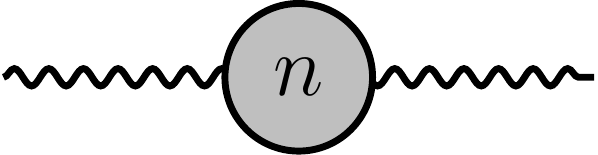}} \,.
\end{align}
We indicate an $n$-loop bubble chain with a gray blob with an index $n$ referring to the number of fermion bubbles in the photon chain.
At the zeroth order in the expansion, only the single-fermion bubble contributes to the beta function, which reads 
\begin{align}
  \label{eq:beta0}
  \beta^{(0)}(K)  &= \frac{2}{3}K^2 \,.
\end{align}
The first order is given by the diagrams
\begin{align}
 \label{eq:beta1-diags}
 \parbox{2.8cm}{\includegraphics[scale=.27]{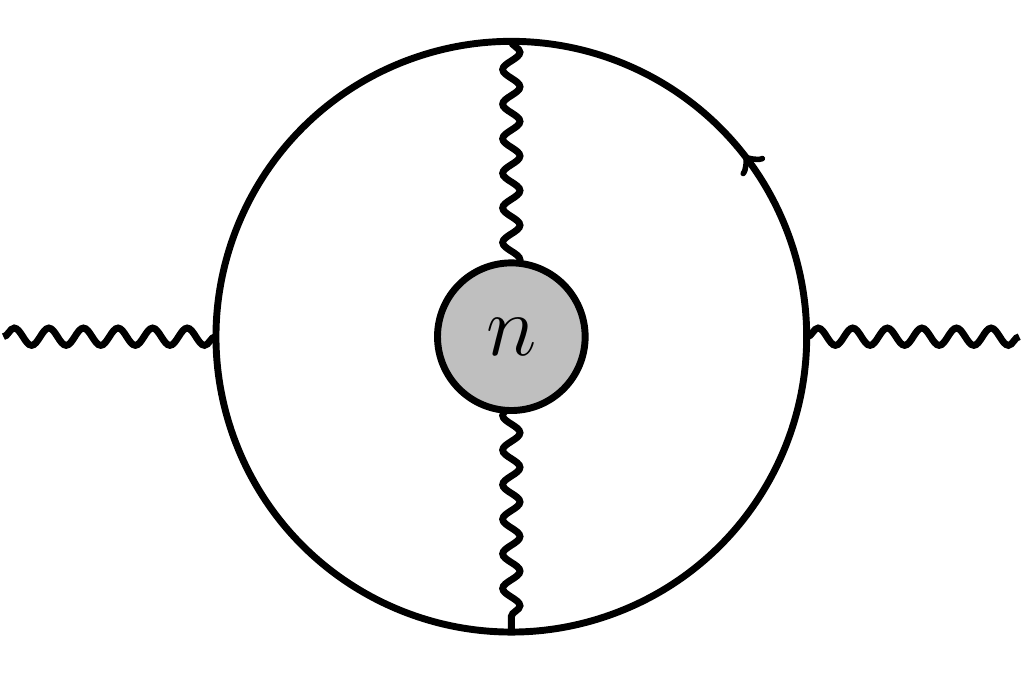}}
 +2 \,\parbox{2.8cm}{\includegraphics[scale=.27]{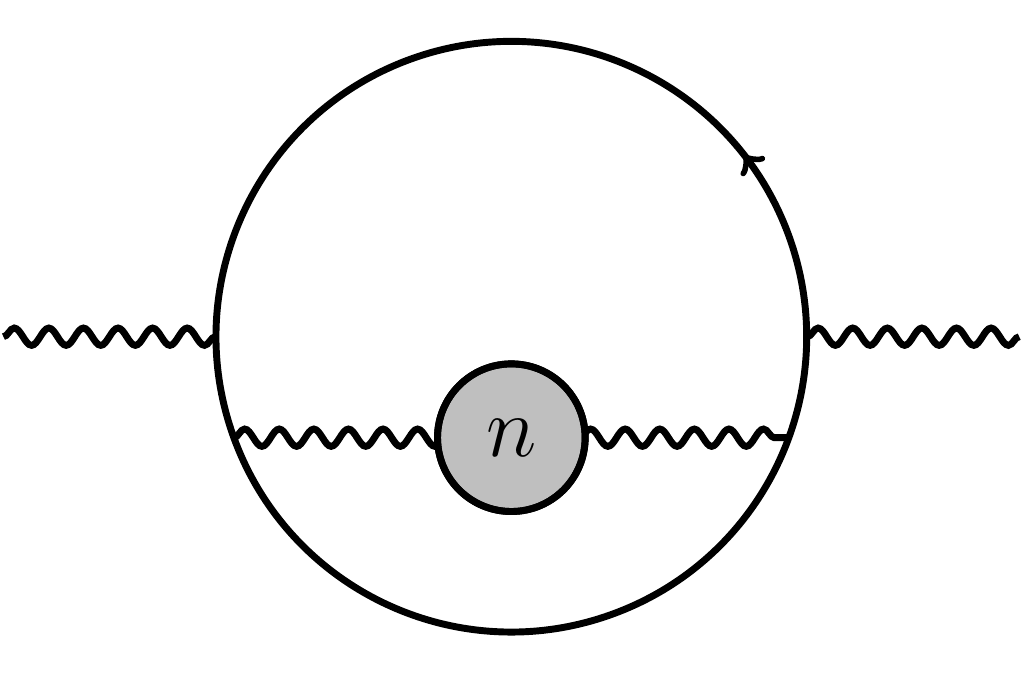}}
  \equiv \parbox{1.6cm}{\includegraphics[scale=.27]{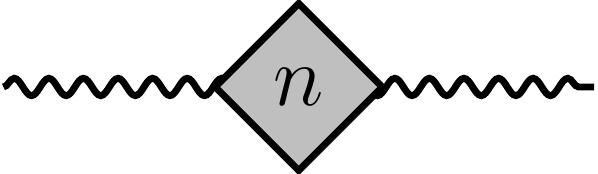}}\,.
\end{align}
We indicate the sum of these diagrams by a gray square labeled by the same $n$. 
The diagrams in \eqref{eq:beta1-diags} were computed for the first time in \cite{PalanquesMestre:1983zy}.
The analogous contribution for QCD was computed in \cite{Gracey:1996he,Holdom:2010qs}, see \cite{Gracey:2018ame} for a review.
Since the diagrams contain only a single bubble chain, the resulting beta function can be resummed and expressed by a closed integral representation
\begin{align}
   \label{eq:beta1}
  \beta^{(1)}(K)  &=  \frac{K^2}{2}  \int_0^K \!\!  \mathrm d x\, F(x)\, . 
\end{align}
The integrand function is given by
\begin{align}
   \label{eq:integrand}
  F(x) &=-
   \frac{(x+3) (x - \frac92) ( x - \frac32) \sin\!\left(\frac{\pi  x}{3}\right) 
  \Gamma\!\left(\frac{5}{2}-\frac{x}{3} \right)}{27\cdot 2^{\frac{2 x}{3}-5} \pi ^{\frac32} (x-3) x \,\Gamma\!\left(3-\frac{x}{3}\right)}   \,. 
\end{align}
This beta function has logarithmic branch cuts at $K_n = \frac{15}2 + 3 n$ for $n\geq0$.
The leading behavior with which it approaches the radius of convergence is
\begin{align}
\label{eq:log}
\beta^{(1)}(K)\sim &\frac{14 K^2}{45\pi^2 }\ln\!\left(\frac{15}{2}-K\right)+\dots \,, \quad 
K\to \frac{15}{2} \,.
\end{align}
Thus, the behavior near the first branch cut is {\it negative}, which allows the $1/N_f$ contribution, \eqref{eq:beta1}, to cancel the leading contribution, \eqref{eq:beta0}.
This leads to a zero in the beta function at this order in $1/N_f$, which has triggered speculations about the existence of a UV fixed point.
Similarly in QCD, a negative logarithmic branch cut at $K=3$ allows for a zero in the beta function.
The potential fixed point in QED has a diverging fermion anomalous mass dimension \cite{Antipin:2017ebo} and is thus considered unphysical.
At the analogous fixed point in QCD, the fermion anomalous mass dimension is instead vanishing \cite{Antipin:2017ebo},
but the fixed point suffers from glueball operators with diverging anomalous dimensions \cite{Ryttov:2019aux}.
In Ref.~\cite{Ryttov:2019aux}, the authors argued that this fact can be interpreted as an operator decoupling and thus the interacting fixed point might still be physical. 
The potential existence of the fixed point has triggered already many phenomenological studies \cite{Mann:2017wzh,Molinaro:2018kjz,Sannino:2019sch}.
The viability of the fixed points has been studied on the lattice \cite{Leino:2019qwk} and with critical point methods \cite{Alanne:2019meg,Alanne:2019vuk}.

Here we go beyond the state of the art by computing part of the full beta function at orders $1/N_f^2$ and $1/N_f^3$.
We are interested in whether new singularities can appear at this order that could shrink the overall radius of convergence.
The complete knowledge of the full beta function at these orders would be ideal to test the physical nature of the potential fixed points. Given the complexity of the task, we focus here on QED and determine  the contributions coming from diagrams of the nested type (to be defined in the next section) to the $1/N_f^2$ and  $1/N_f^3$ order. We will see that these contributions alone show rather interesting features.

\begin{figure*}[t]
\includegraphics[width=.45\linewidth]{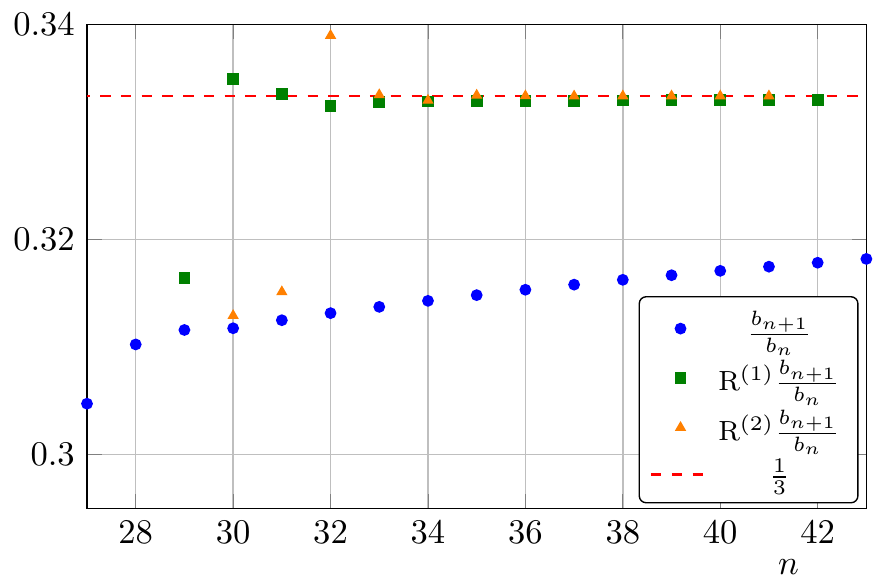} \hfill
\includegraphics[width=.45\linewidth]{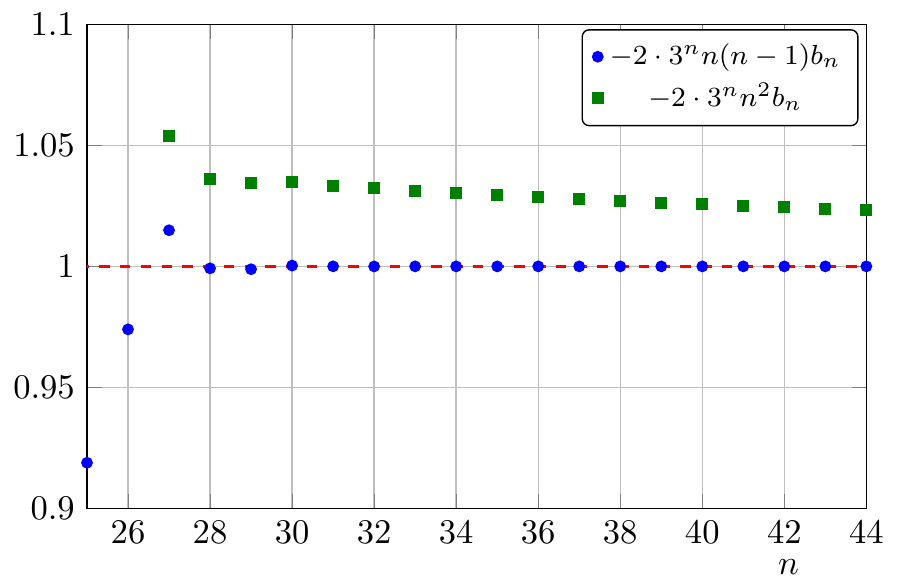} 
\caption{Left: The ratio test applied to the expansion coefficients of $\beta_{\rm nested}^{(2)}(K)$ reveals that the radius of convergence is $K=3$. A Richardson extrapolation is used to accelerate the convergence of the series.
Right: The prefactor of the leading large-order behavior is determined to be $-\frac{1}{2n (n-1) 3^n}$, which leads to a much faster convergence than $-\frac{1}{2n^2 3^n}$. See App.~\ref{app:Darboux}.
}
\label{fig:richardson}
\end{figure*}

\section{Nested diagrams}
\label{sec:CND}
In \cite{Dondi:2019ivp}, we have laid the foundations to access  crucial information regarding the singular structure of the beta function of the theory by knowing finitely many coefficients of the perturbative series in the coupling $K$.  We showed that to extract the precise radius of convergence and uncover the first singularity in $K$, roughly the first thirty coefficients of the perturbative expansions are needed. Clearly, the task to extract so many coefficients becomes progressively more demanding when going beyond the leading result. Therefore, although, the procedure can, in principle, be applied to the full $1/N_f^2$ and $1/N_f^3$ beta function, already to $\mathcal{O}(1/N_f^2)$ it requires to determine four-loop diagrams and non-planar three-loop diagrams, for which the master integrals with dressed propagators are not known, see \autoref{fig:beta2-diags} in App.~\ref{app:renormalization} for the full set of diagrams.

Fortunately there is one closed set of diagrams, the nested ones, which is gauge and RG-scale independent, and which can be tackled. The associated Feynman diagrams are obtained iterating the $1/N_f$ topologies and are given by
\begin{align}
  \label{eq:beta2-diags}
 \parbox{2.8cm}{\includegraphics[scale=.27]{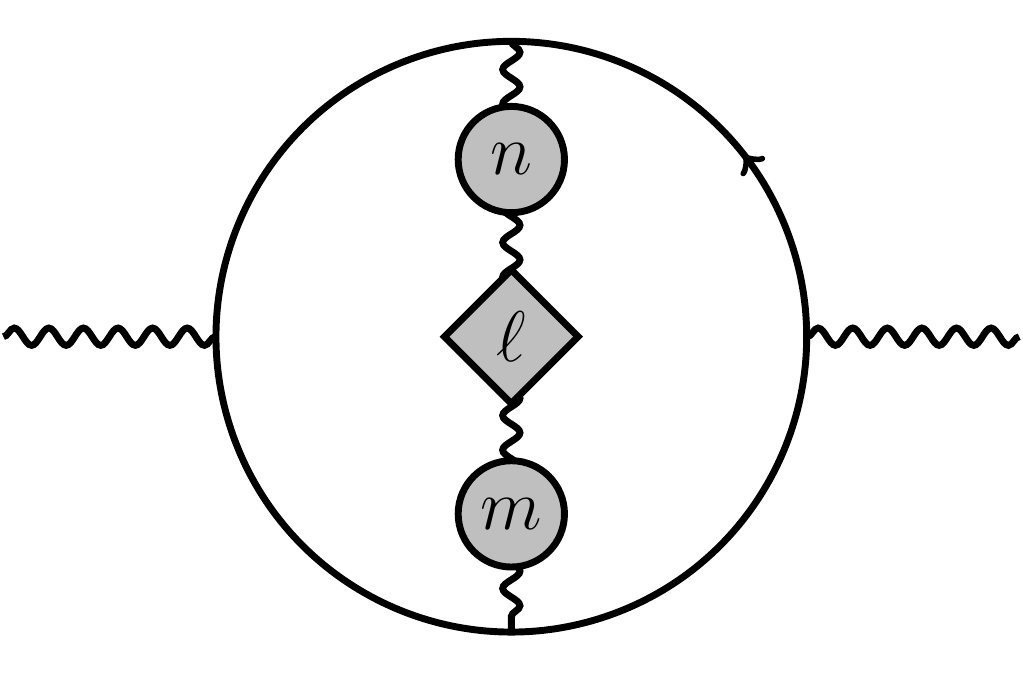}}
 +2 \,\parbox{2.8cm}{\includegraphics[scale=.27]{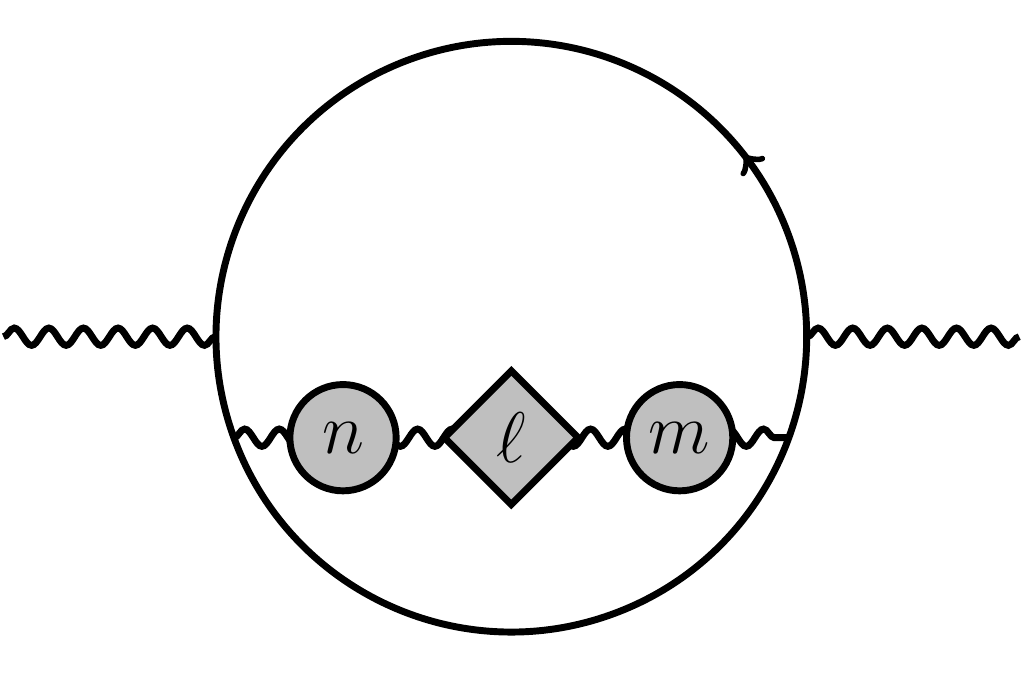}}
  \equiv \parbox{1.6cm}{\includegraphics[scale=.27]{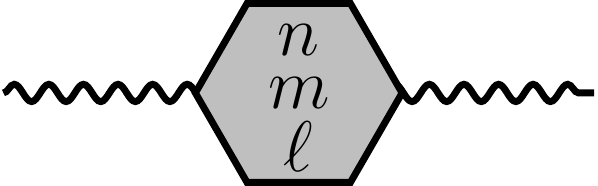}}\,.
\end{align}
We represent the sum of these contributions with a gray hexagon and three indices labeling the number of fermion bubbles on each photon propagator.
The full amplitude from these diagrams is given in \eqref{eq:beta2-amplitude} of App.~\ref{app:renormalization} in terms of discrete convolutions of the $1/N_f$ amplitude.
There is an additional counterterm contribution that stems from inserting a $1/N_f$ counterterm on the photon line of the $1/N_f$ diagrams in \eqref{eq:beta1-diags}. The explicit form of the counterterm contribution is given in \eqref{eq:Zn}. The sum of all these contributions is gauge and RG scale independent, which we verified by explicit computation. We computed these contributions to the beta function separately up to $K^{44}$.
The coefficients are listed in App.~\ref{app:beta-function}.

 At $\mathcal{O}(1/N_f^3)$ the contributing diagrams are given by
\begin{align}
  \label{eq:beta3-diags}
 \parbox{3.1cm}{\includegraphics[scale=.3]{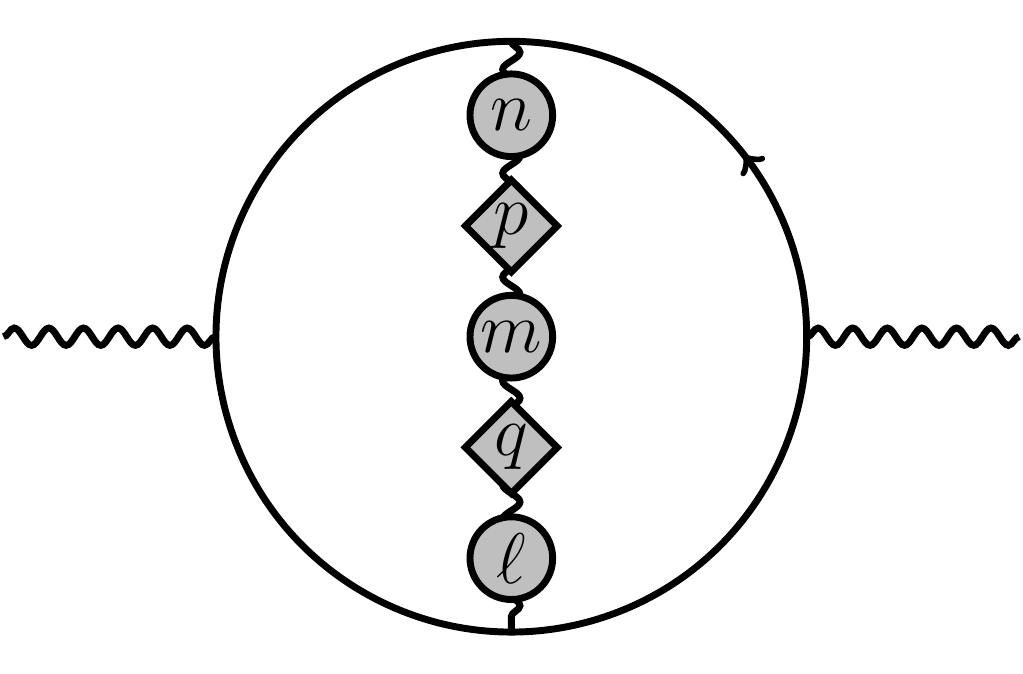}}
 &+2 \,\parbox{3.1cm}{\includegraphics[scale=.3]{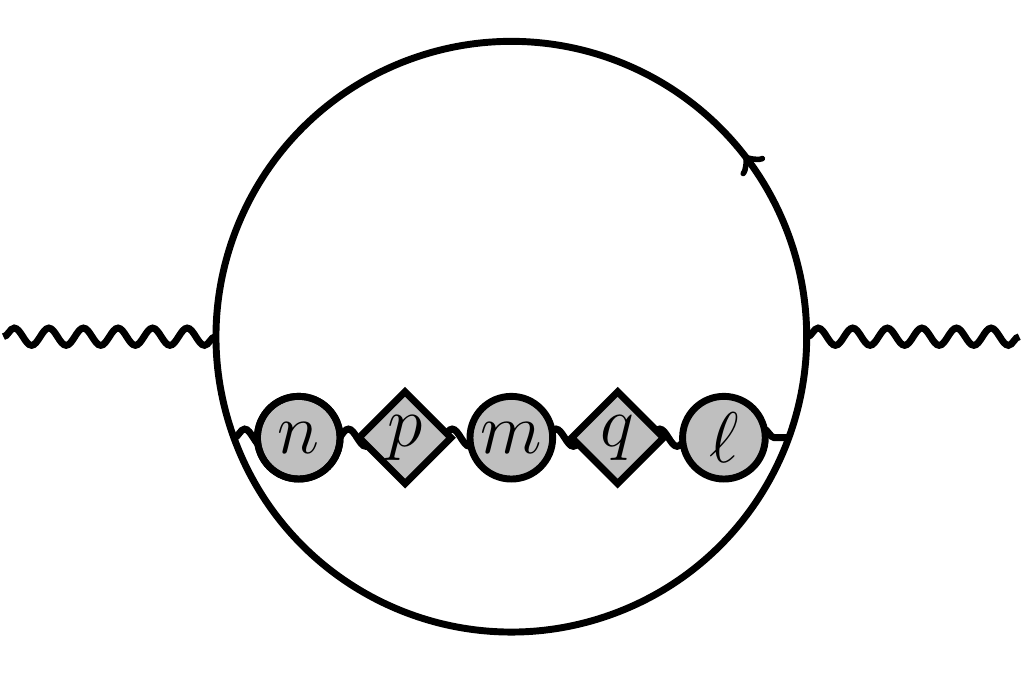}} \notag \\
 +\,\parbox{3.1cm}{\includegraphics[scale=.3]{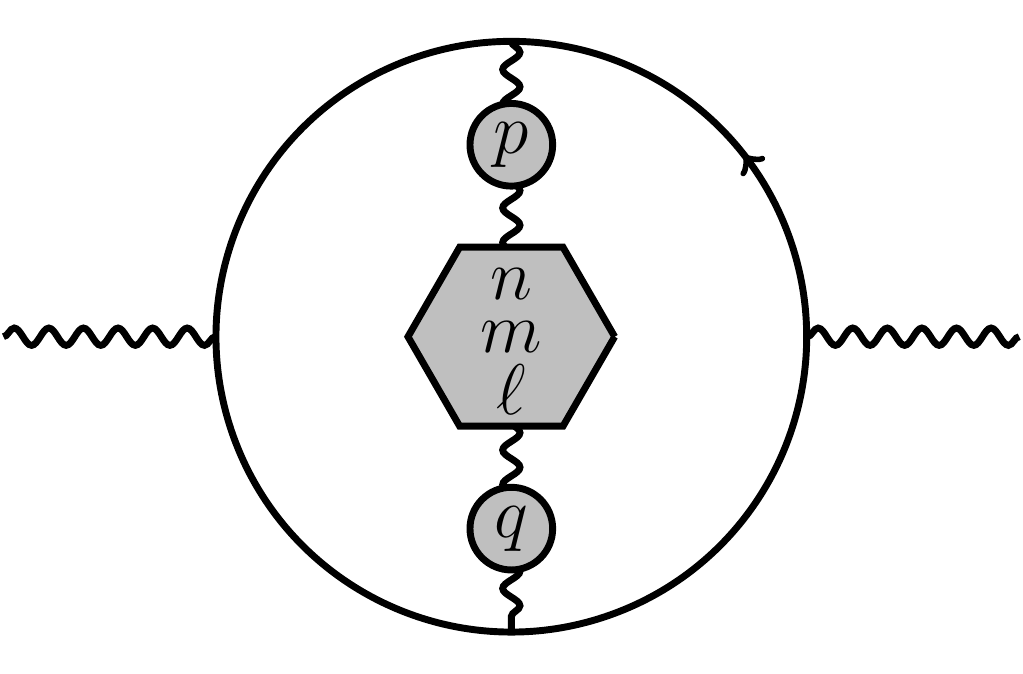}}
&+2\, \parbox{3.1cm}{\includegraphics[scale=.3]{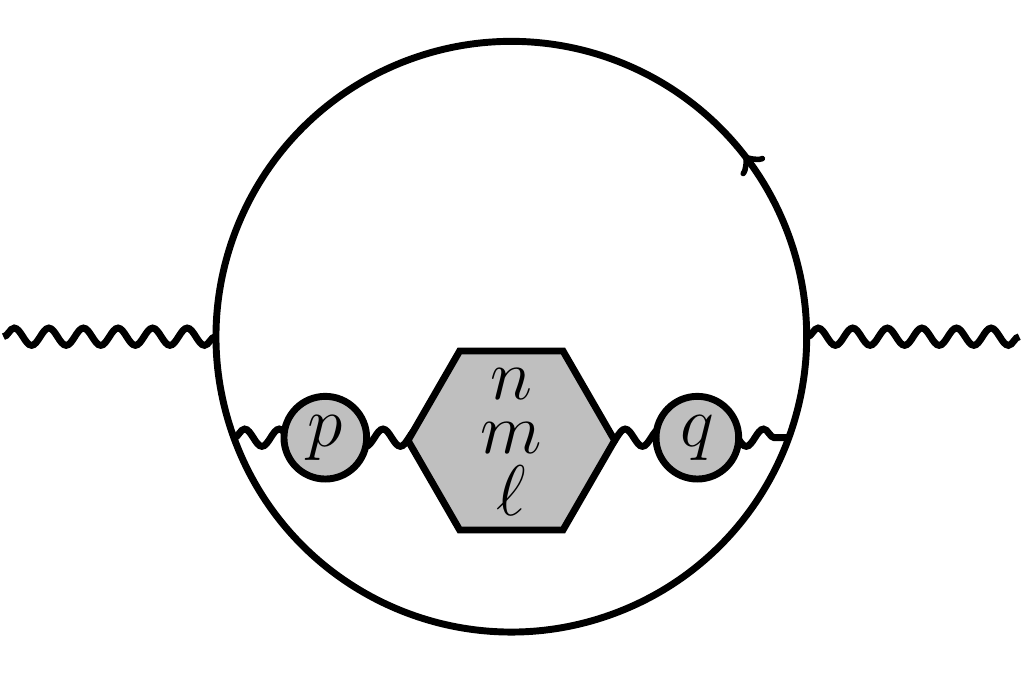}}\,.
\end{align}
The full amplitudes for these diagrams are now given in  \eqref{eq:beta3-amplitude1} and \eqref{eq:beta3-amplitude2} of App.~\ref{app:renormalization}.
We determined these contributions to the beta function up to  $K^{32}$ and report the coefficients in App.~\ref{app:beta-function}, together with a comparison with the total 5-loop result from \cite{Baikov:2012zm, Herzog:2017ohr}.

Determining the coefficients from the diagrams in \eqref{eq:beta2-diags} and \eqref{eq:beta3-diags} required a significant computational effort. We used the Mathematica package \textit{FeynCalc} \cite{Shtabovenko:2016sxi} to contract the diagrams and standard multi-loop techniques to evaluate them, see, for example, \cite{Broadhurst:1996ur, Bierenbaum:2003ud, Grozin:2003ak, Grozin:2012xi}. The most computation power is needed to numerically extract the divergent and finite part at each loop order. The Mathematica package \textit{NumExp} \cite{Huang:2012qz}, which numerically expands hypergeometric functions, turned out to be very useful in this context. Hypergeometric functions naturally appear from the evaluation of loop integrals, see App.~\ref{app:renormalization}.

In general, the $\beta$-function coefficients $\beta^{(n)}$ are dependent on the used scheme. The lowest order $\beta^{(0)}$ is scheme-independent since it is a one-loop result. The order $\beta^{(1)}$ is scheme-independent as well if the functional relation between the couplings in the two schemes does not involve $N_f$ \cite{Shrock:2013cca}. In turn, the higher coefficients $\beta^{(n>1)}$ are scheme-dependent functions of the coupling $K$. The singularity structure of these functions, and thus the large-order behaviour of their series in $K$, is scheme-invariant as long as the functional relation between the couplings in the two schemes is sufficiently regular. We employ dimensional regularisation in the minimal subtraction scheme.

We are now ready to analyze the large-order behavior of the expansion coefficients of $\beta_\text{nested}^{(2)}(K)$, in order to extract physical information concerning possible physical singularities. We apply ratio tests, and  Darboux's theorem \cite{Fisher, gaunt1974asymptotic, Henrici},  as well as Pad\'e methods to access information about the leading singular structure of the associated beta function. To keep the presentation light we report the details of the methods in App.~\ref{app:Darboux} and App.~\ref{app:pade}. 

\subsection{Leading singularity of \texorpdfstring{$\beta^{(2)}_\text{nested}$}{beta2nested}}
To extract the leading singularity we must first and foremost demonstrate that the number of terms at our disposal is sufficient to see convergence. This is performed by running the ratio test $b_{n+1}/b_{n}$, with $b_n$ the coefficients of the  $\beta^{(2)}_\text{nested}$ series. For sufficiently large $n$ the ratio approaches the inverse radius of convergence.
We report the results in the left panel in \autoref{fig:richardson} and give the detailed analysis in App.~\ref{app:Darboux}. To accelerate the convergence we further employed Richardson extrapolation. From the left plot in \autoref{fig:richardson}, one learns that about thirty coefficients are sufficient to approach the convergence of the series.

One learns that the radius of convergence is $K=3$. From the right plot in \autoref{fig:richardson}, we further learn that in total the coefficients have leading decay $b_n\sim -1/(2\cdot 3^n n^2)$. This allows us to subtract from each $b_n$ its leading large $n$ contribution, allowing us to determine its sub-leading large  $n$ behavior which is discovered to go as $-1/(2\cdot 3^n n^3)$. This trend repeats after each subtraction and therefore we can determine the large-order behavior of the nested QED beta function coefficients at  $\mathcal{O}(1/N_f^2)$ to be
\begin{align}
\label{eq:large-order}
b_n &\sim -\frac12 \frac1{3^n} \left( \frac1{n^2} + \frac1{n^3} + \dots \right)
= -\frac12 \frac1{3^n} \frac1{n(n-1)}\,.
\end{align}
It is interesting that with just 30 expansion coefficients we can clearly distinguish the correct sub-leading large $n$ behavior, as shown in the right plot in \autoref{fig:richardson}. Note that the behavior in \eqref{eq:large-order} is the natural sub-leading behavior for a logarithmic singularity: see App.~\ref{app:Darboux}. 
This exact large $n$ order behavior determines the nature of the first singularity once we resum the series 
\begin{align}
-\frac12\sum_{n=4}^\infty \frac1{n(n-1)} \frac{K^n}{3^n}  &= \frac16 (K-3) \ln\!\left(1-\frac{K}{3}\right) +\text{regular} \,.
\label{eq:branch-cut}
\end{align}
This implies that in the vicinity of the leading singularity at $K=3$, 
\begin{align}
\label{eq:beta2tilde}
 \beta^{(2)}_\text{nested}(K) \sim -\frac12 \left(1-\frac{K}{3}\right) \ln\!\left(1-\frac{K}{3}\right) +\dots\,, \quad K\to 3 \,.
\end{align}
where the sub-leading terms are analytic at $K=3$.

\begin{figure}[t]
\includegraphics[width=\linewidth]{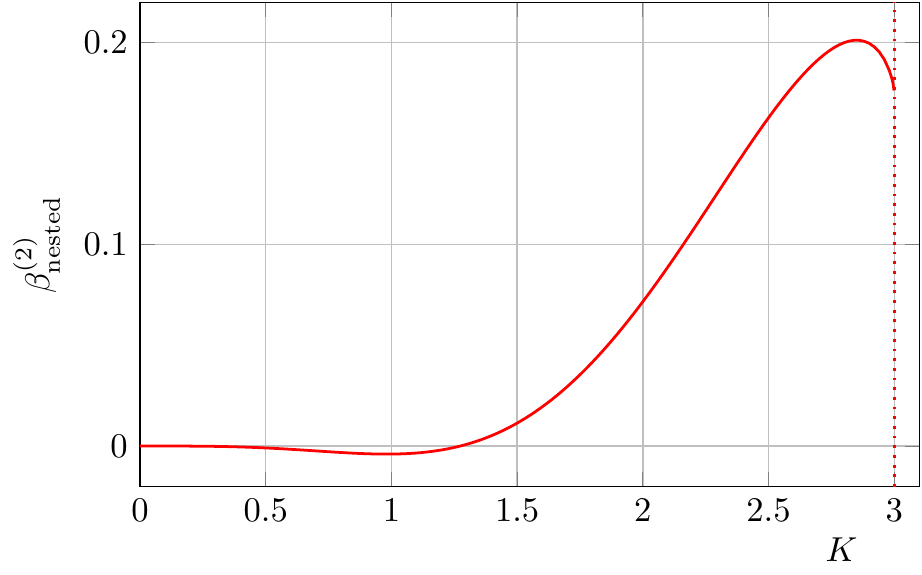}
\caption{Nested QED beta function at $\mathcal{O}(1/N_f^2)$. At $K=3$ the beta function is finite, but it has a logarithmic branch cut.}
\label{fig:nested-beta}
\end{figure}

This is a remarkable result for a number of reasons: i) The nested QED beta function at $\mathcal{O}(1/N_f^2)$ has a logarithmic branch cut at $K=3$ while remaining finite there. ii) Once the large order behavior is subtracted the remaining contribution at $K=3$ is regular. iii) The singularity occurs at a value of $K$ which is smaller than the leading singularity of QED occurring for $K=15/2$ at the first order in $1/N_f$. iv) The singularity occurs at the same value as the leading one for QCD. 

It is worth mentioning that because of the simple structure of the function multiplying the logarithmic singularity in \eqref{eq:branch-cut}, it is possible to confirm this behavior by analyzing the series obtained by a second derivative with respect to $K$ of the nested beta function. This is so because the logarithmic singularity turns into a simple pole that it is more easily accessed by the test. In fact, in this case, the onset of the converge occurs already for $\sim K^{28}$ as detailed in \autoref{fig:derivatives} of App.~\ref{app:Darboux}.
 
We can now use Pad\'e approximants to deduce the full form of the nested beta function up to the singularity, which is depicted in \autoref{fig:nested-beta}. More information about the Pad\'e analysis is provided in App.~\ref{app:pade}. 

\begin{figure}[t]
\includegraphics[width=\linewidth]{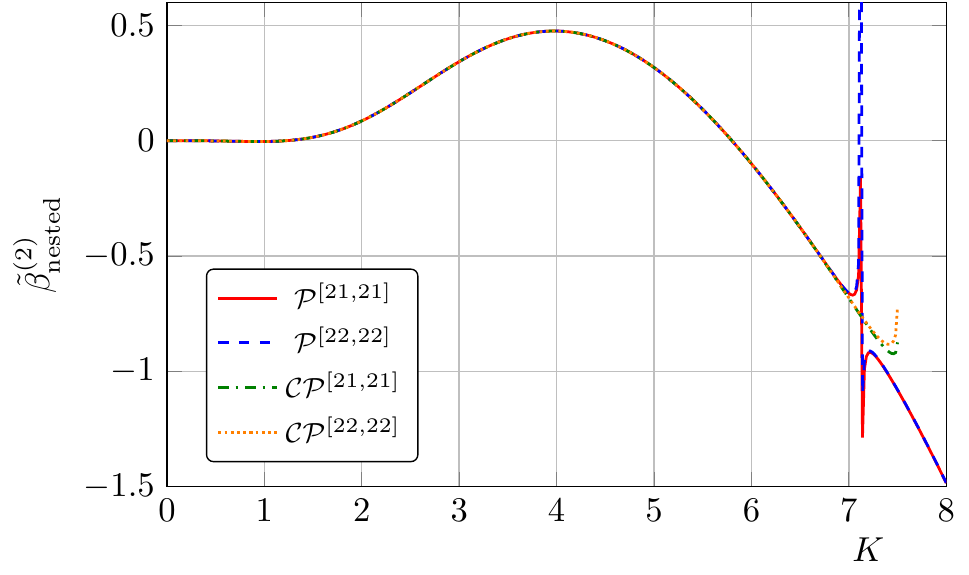}
\caption{(Conformal) Pad\'e approximants of the nested QED beta function at $\mathcal{O}(1/N_f^2)$ with the branch cut  subtracted, see \eqref{eq:beta2tilde}.
The Pad\'e approximants seem to hint towards a pole or a branch-cut at $K=\frac{15}{2}$.
}
\label{fig:pade-1N2}
\end{figure}

\subsection{Sub-leading singularity of \texorpdfstring{$\beta^{(2)}_\text{nested}$}{beta2nested}}
We now turn to the sub-leading singular behavior of $\beta^{(2)}_\text{nested}$.
To that end we subtract the leading logarithmic branch cut behavior from the nested beta function
\begin{align}
\label{eq:beta2tilde}
\tilde \beta^{(2)}_\text{nested} = \beta^{(2)}_\text{nested} +\frac12\sum_{n=4}^\infty \frac1{n(n-1)} \frac{K^n}{3^n}  \,.
\end{align}
With the coefficients at hand no further singular behavior is revealed by the ratio test. 
From the naive expectation that the next singularity arises at $K=\frac{15}2$, which is the point where the leading order full beta function is singular, we estimate that roughly 55 coefficients would be needed, which goes beyond the scope and resources of this investigation. 

Since the ratio test is not sufficiently precise to display any sub-leading singular behavior, we use Pad\'e methods instead.
The three highest Pad\'e approximants are displayed in \autoref{fig:pade-1N2}.
They converge well up to $K\approx 7$ and thus we know that there is no singular behavior for $K<7$.
For $K>7$, some approximants show singularities and we suspect to find the next singular behavior at $K=\frac{15}2$, since this was the location of the logarithmic branch cut at $1/N_f$.

Having a hint for the location of the next singularity, we now employ the conformal Pad\'e method which is expected to be more accurate in this case. The conformal Pad\'e takes as input the location of the singularity and maps the series to the unit disc. After the conformal transformation, one re-expands the function and applies the standard  Pad\'e method and, in the end, one inverts back the conformal transformation. The Pad\'e-Conformal method is described in detail in App.~\ref{app:pade}. In \cite{Costin:2019xql}, the improvement by a suitable conformal transformation was studied on the example of the Painlev\'e I equation.  We tested the conformal Pad\'e method for the leading singularity at $1/N_f^2$ and it indeed gives improved results, for details see App.~\ref{app:pade}, and \autoref{fig:pade-1N2} and \autoref{fig:CPade-1N2}.

For the sub-leading singularity, the conformal Pad\'e results are displayed in \autoref{fig:pade-1N2}. We therefore feel confident to have captured both the leading branch-cut singularity at $K=3$, and the sub-leading one occurring at $K=15/2$. However, for the latter the available data (i.e.\ the available expansion coefficients) is not sufficient to determine the precise nature of this sub-leading singularity. 

\begin{figure}[t]
\includegraphics[width=\linewidth]{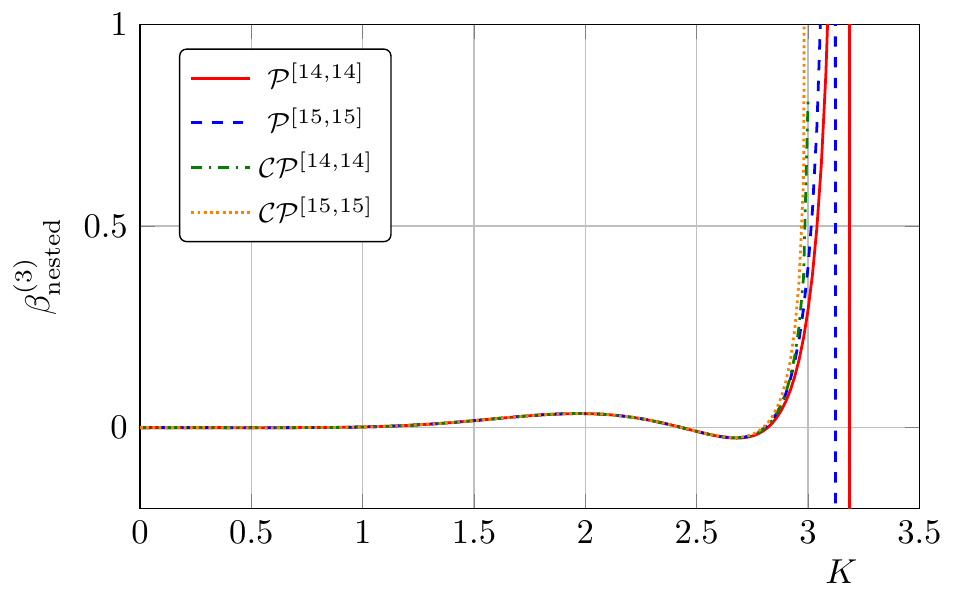}
\caption{(Conformal) Pad\'e approximants of the nested QED beta function at $\mathcal{O}(1/N_f^3)$.
The Pad\'e approximants seem to hint towards a pole or a branch-cut at $K=3$.}
\label{fig:pade-1N3}
\end{figure}

\subsection{Leading singular behavior of \texorpdfstring{$\beta^{(3)}_\text{nested}$}{beta3nested}}
Here we use directly the Pad\'e method to infer the location of the first singularity given the fewer computable coefficients than needed for the ratio test. The highest-order Pad\'e approximants are displayed in \autoref{fig:pade-1N3} showing convergence up to $K\approx 2.7$ and hence we can exclude any singular behavior in this range. It is reasonable  to expect the first singularity to occur for $K=3$, since all Pad\'e approximants have a pole shortly after $K=3$.
We therefore apply the conformal Pad\'e method (see App.~\ref{app:pade}) to get a more accurate representation of the associated beta function up to the singularity, and plot it in \autoref{fig:pade-1N3}.
 
\begin{figure}[t]
	\includegraphics[width=\linewidth]{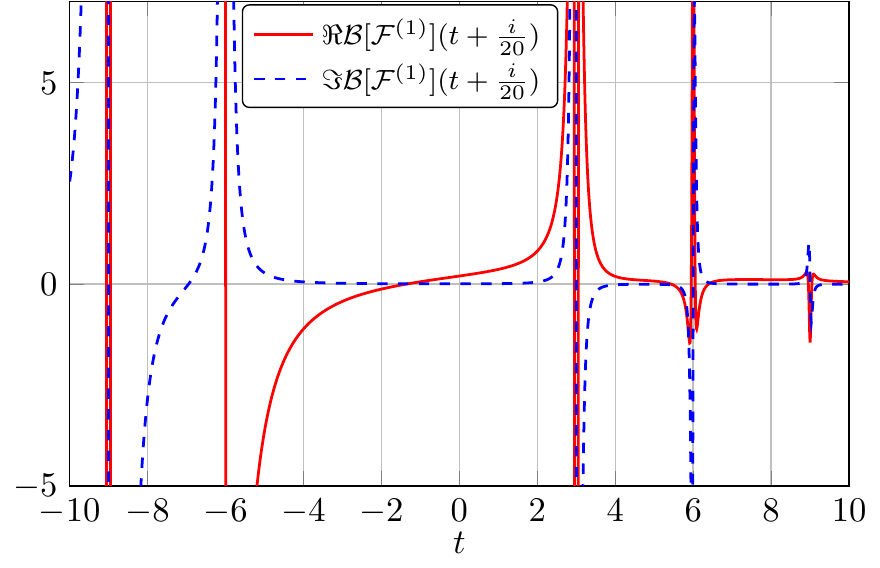}
	\caption{Real and imaginary parts of the Borel transform of the finite part of the renormalized photon two-point function at order $1/N_f$.
		This Borel transform is obtained from the analytic formula \eqref{eq:borel1} in App.~\ref{app:finite-parts}.
		We see UV renormalon singularities at $t=3k$ with $k=1, 2, \ldots, \infty$, and IR renormalon singularities at $t=-3k$ with $k=2, 3,\ldots, \infty$.
	}
	\label{fig:Borel-finite-part}
\end{figure}

\begin{figure}[b]
	\includegraphics[width=\linewidth]{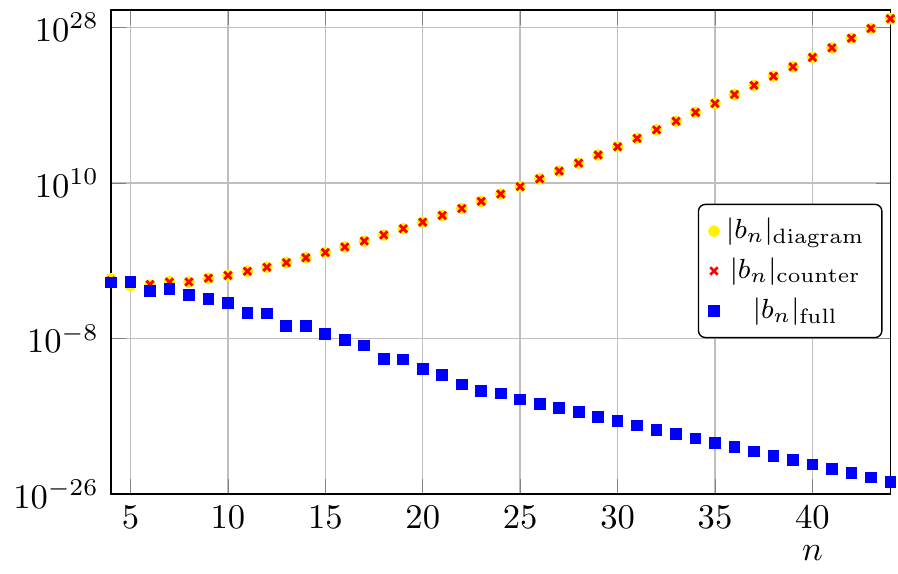}
	\caption{The $1/N_f^2$ coefficients of the $1/\epsilon$ part of the nested diagrams and corresponding counter term are each factorially divergent.
	As these are not separately RG-independent we choose $\mu^2 = - p^2/4\pi$, $p^2$ being the external momentum.
	Their sum produces a RG-independent convergent series according to the large order behavior given in \eqref{eq:large-order}.
	}
	\label{fig:factorial-divergence}
\end{figure}

\section{Finite Parts and Renormalons}
\label{sec:renormalons}
So far, we have discussed the divergent part of the photon two-point correlation function  because it is directly related to the beta function of the theory. The finite part, however, also has  an interesting story to tell, the renormalon story \cite{Beneke:1998ui}. Indeed, the class of diagrams contributing to the $1/N_f$ correlator are the ones that were originally considered as a renormalon source. Renormalons emerge as singularities of the Borel transform of the finite part of the correlation function and produce a factorially growing series not associated with diagram proliferation.
Resummation methods for the finite parts in QED at order $1/N_f$ have been explored in \cite{PalanquesMestre:1985ty} and shown not to be Borel summable (in the massless fermion case), due to poles located at $3k$, with $k=1,\ldots, \infty$, along the positive real Borel axis. This corresponds to a leading factorial growth $n!/3^n$  of the coefficients of the original finite part. The finite parts at higher orders of the $1/N_f$ expansion are analyzed in App.~\ref{app:finite-parts}. The $1/N_f$ result for the Borel transform of the finite part of the photon two-point function is derived in \eqref{eq:borel1}, and is plotted in \autoref{fig:Borel-finite-part}, just above the real Borel $t$ axis. This plot clearly indicates the appearance of singularities at $t=3, 6, 9, ...$ on the positive Borel axis, and singularities on the negative Borel axis at $t=-6,-9, ...$. In the minimal subtraction scheme, renormalons affect the finite parts only, and this is confirmed at the $1/N_f$ order.

Remarkably, even with our limited number of expansion coefficients, we observe the same factorial growth and singularity structure at sub-leading orders $1/N_f^2$ and $1/N_f^3$ of the divergent part of the nested diagrams. See \autoref{fig:factorial-divergence} for the rate of growth of the expansion coefficients in the $1/N_f^2$ case. The coefficients for the nested diagrams and the corresponding counterterm both grow factorially fast, but with the same rate and opposite signs, in such a way that the factorial growth cancels, leaving coefficients of a rapidly convergent expansion, as shown in \eqref{eq:large-order}. 

\begin{figure}[t]
	\includegraphics[width=\linewidth]{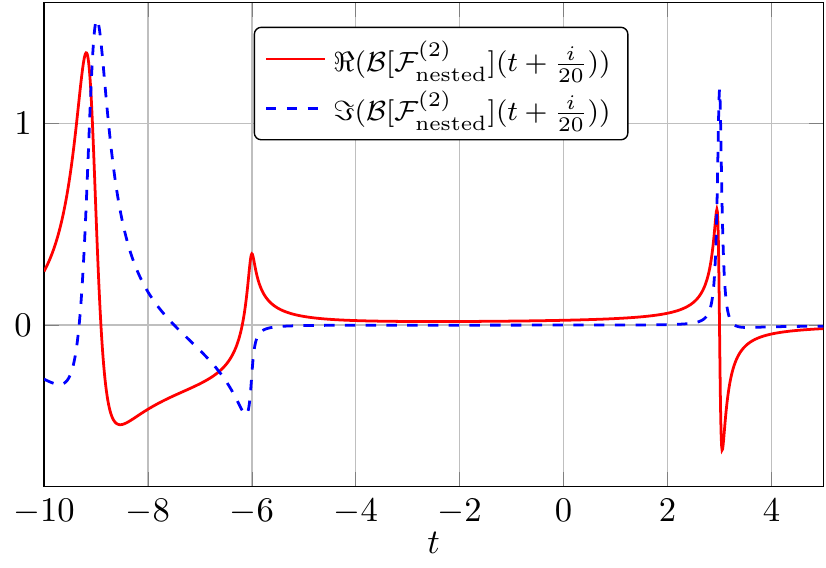}
	\caption{Real and imaginary parts of the Borel transform of the finite part of the renormalized photon two-point function at order $1/N_f^2$.
	This Borel transform has been reconstructed from the finite number of expansion coefficients using the Pad\'e-Conformal method described in App.~\ref{app:pade}.
	We see a UV renormalon singularity at $t=+3$, an IR renormalon singularity at $t=-6$, and a hint of a further  IR renormalon singularity at $t=-9$.
	}
	\label{fig:Borel-finite-part2}
\end{figure}

\begin{figure}[t]
	\includegraphics[width=\linewidth]{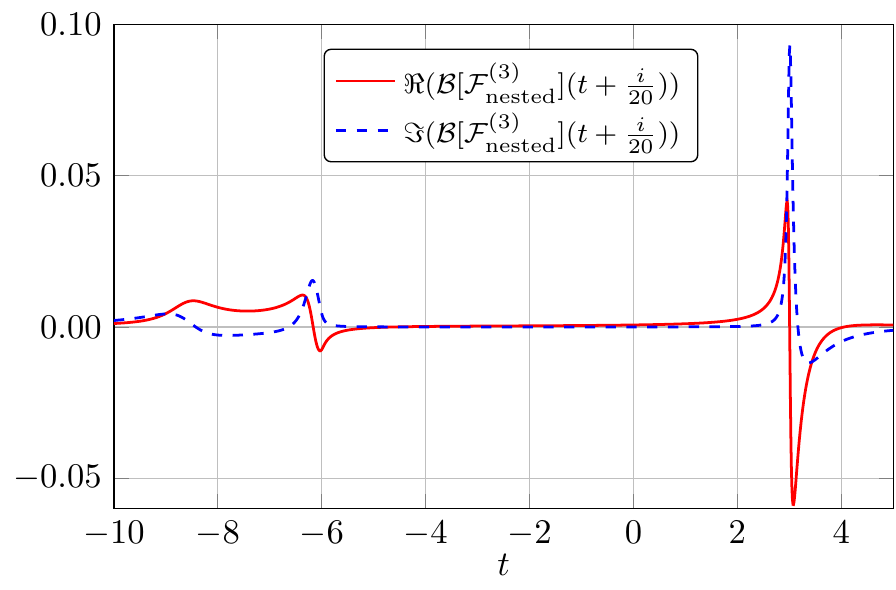}
	\caption{Real and imaginary parts of the Borel transform of the finite part of the renormalized photon two-point function at order $1/N_f^3$. 
	This Borel transform has been reconstructed from the finite number of expansion coefficients using the Pad\'e-Conformal method described in App.~\ref{app:pade}.
	We see a UV renormalon singularity at $t=+3$, an IR renormalon singularity at $t=-6$, and a hint of a further IR renormalon singularity at $t=-9$.
	}
	\label{fig:Borel-finite-part3}
\end{figure}

Furthermore, in \autoref{fig:Borel-finite-part2} and \autoref{fig:Borel-finite-part3} we plot the real parts of the Borel transform of the
finite parts at orders $1/N_f^2$ and $1/N_f^3$, respectively, as shown in \eqref{eq:finite_parts_def}. These plots should be compared with the $1/N_f$ result in \autoref{fig:Borel-finite-part}. At orders $1/N_f^2$ and $1/N_f^3$ we do not have the luxury of closed-form expressions, but with the limited number of expansion coefficients (see App.~\ref{app:beta-function}) our Borel transforms, after conformal mapping and Pad\'e approximation (see App.~\ref{app:pade}), clearly reveal singularities at $t=+3$ and $t=-6$, with strong indications of a further singularity at $t=-9$. With more coefficients, one would be able to resolve even more Borel singularities. Note that without the conformal map, the Pad\'e approximation to the Borel transform cannot see any physical singularities beyond the leading ones, because Pad\'e tries to represent the leading branch cut with an array of poles and zeros, which have no physical content beyond a crude representation of the cut, and these unphysical poles therefore obscure further physical singularities. On the other hand, the Pad\'e approximation to the conformally-mapped expansion, as described in App.~\ref{app:pade}, does not place unphysical singularities on the cut \cite{Costin:2019xql,Costin:2020hwg}, so higher physical singularities can be seen. 
These results suggest a relation between the leading order renormalon factorial growth and the singularities in the divergent part of the nested diagrams. This QED Borel structure suggests singularities on the positive real axis associated with UV renormalons, and singularities on the negative real axis associated with IR renormalons \cite{Beneke:1998ui}.

\section{Conclusions and outlook}
\label{sec:conclusions}
In this paper we have determined the contribution to the QED beta function stemming from the gauge and RG-scale independent class of nested diagrams to order $1/N_f^2$ and $1/N_f^3$,  resolving their leading singularity structure. We have shown that:
\begin{itemize}
 \item [1)] The nested beta function at $\mathcal{O}(1/N_f^2)$ has a new logarithmic branch cut at $K=3$, coinciding with the QCD branch cut at $\mathcal{O}(1/N_f)$.
 	The nested beta function is finite at the branch cut.
 \item[2)]  The next singularity of the nested beta function appears  at $K=\frac{15}2$. However, we do not have enough perturbative data to fully characterize its nature. 
  \item[3)] The first singularity of the nested beta function at $\mathcal{O}(1/N_f^3)$  appears  at  $K=3$, but its nature remains to be determined.
 \item[4)] We observed that the factorial growth of the divergent part of the nested diagrams at $\mathcal{O}(1/N_f^2)$ matches the one for the finite part of the leading $1/N_f$ contribution which is related to the renormalons of the theory. An analogous structure  is also seen at   $\mathcal{O}(1/N_f^3)$.
\end{itemize}

An important message from this analysis is that it is indeed feasible, as proposed in \cite{Dondi:2019ivp}, to use a finite number of perturbative expansion terms to probe certain  non-perturbative properties at higher orders of the large $N_f$ expansion; we do not require the closed-form expressions which are available at leading order. This suggests a new strategy for studying  physical properties of large $N_f$ expansions.

\vspace{.3cm}
\noindent {\bf Acknowledgements} 
We would like to thank Anders Thomsen, Tommi Alanne and Simone Blasi for discussions on the finite part resummations.
This work is partially supported by the Danish National Research Foundation grant DNRF:90, and  is based upon work supported by the U.S. Department of Energy, Office of Science, Office of High Energy Physics under Award Number DE-SC0010339.  

\appendix
\begin{widetext}
\section{Renormalization procedure and computation of nested diagrams}
\label{app:renormalization}
In this appendix, we detail the applied renormalization procedure.
We apply dimensional regularization in $d = 4 - \epsilon$ dimensions.
The 1PI photon two-point function is parameterized by
\begin{align}
\Gamma^{(2)}_{\mu\nu}(p) = p^2 \left( \eta_{\mu\nu} - \frac{p_{\mu}p_{\nu}}{p^2} \right) \Pi(K_0, p^2) \,.
\end{align}
We expand the renormalization of the coupling $Z_K=K/K_0$, where $K_0$ is the bare 't Hooft coupling, as well as $\Pi$ in orders of $N_f$
\begin{align}
Z_{K} &= Z_0 + \frac{1}{N_f} Z_1 + \frac{1}{N_f^2} Z_2 + \frac{1}{N_f^3} Z_3 +  \mathcal{O}(1/N_f^4) \,, \\
\Pi(K_0) &= \Pi_0( Z_K^{-1} K ) + \frac{1}{N_f}\Pi_1( Z_K^{-1} K ) + \frac{1}{N_f^2}\Pi_2( Z_K^{-1} K ) + \frac{1}{N_f^3}\Pi_3( Z_K^{-1} K ) +  \mathcal{O}(1/N_f^4)\,.
\end{align}
Here and in the following we suppress the momentum dependence of $\Pi$ to improve the readability. Each $Z_n$ can be written as a series in $1/\epsilon$
\begin{align}
\label{eq:Zeps}
Z_n = \sum_{i} \frac{Z_n^{(i)}}{\epsilon^i} \,.
\end{align}
The simple pole in $\epsilon$ given by $Z_n^{(1)} $ determines the beta function at each order in $1/N_f$.
The latter is given by 
\begin{align}
\beta^{(n)} = \left[ 1 - K \frac{\partial}{\partial K} \right] Z_n^{(1)} K = - K^2 \frac{\partial}{\partial K} Z_n^{(1)} \,.
\end{align}
We use a minimal subtraction scheme and thus the renormalization condition is
\begin{align}
\text{div}\{ Z_K [1 - \Pi (K_0) ]  \} = 0 \,.
\end{align}
Here, the operator div extracts the parts that are divergent in the limit $\epsilon \to0$.  
We expand this equation in orders of $1/N_f$. From this, also using the fact that $\Pi_0$ is linear in the bare coupling, we obtain
\begin{align}
\label{eq:Zn}
Z_0 ={}& 1 + \text{div}\!\left\{\Pi_0 (K)\right\} ,  \qquad\qquad
Z_1 = \text{div}\!\left\{Z_0 \Pi_1( Z_0^{-1} K)\right\}  ,  \notag \\
Z_2 ={}& \text{div}\!\left\{Z_0 \Pi_2( Z_0^{-1} K)\right\} + \text{div}\!\left\{ Z_1 \left[ 1 - K_0 \frac{\partial}{\partial K_0} \right] \Pi_1(K_0) \right\}_{K_0 = Z_0^{-1} K} \,, \notag \\
Z_3 ={}& \text{div}\!\left\{Z_0 \Pi_3( Z_0^{-1} K)\right\}  + \text{div}\!\left\{ Z_1 \left[ 1 - K_0 \frac{\partial}{\partial K_0} \right] \Pi_2(K_0) \right\}_{K_0 = Z_0^{-1} K} \notag\\
&+ \text{div}\!\left\{ \left( Z_2 - K \frac{Z_2}{Z_0} \frac{\partial}{\partial K_0} + \frac{K^2}{2} \frac{Z_1^2}{Z_0^3} \frac{\partial^2}{\partial K_0^2} \right) \Pi_1(K_0)\right\}_{K_0 = Z_0^{-1} K} \,.
\end{align}
Here, $\Pi_0$ is precisely the single-fermion bubble and thus $Z_0 = 1 - \frac{2 K}{3 \epsilon}$.
$\Pi_1$ is given by the diagrams displayed in \eqref{eq:beta1-diags}.
In $Z_2$, the first term contains the factor $\Pi_2$, which is precisely the diagrams displayed in \autoref{fig:beta2-diags}.
The second term can be viewed as the $1/N_f$ diagrams \eqref{eq:beta1-diags} with a $1/N_f$ counter term insertion.
In $Z_3$, the first term is again given by $1/N_f^3$ diagrams, while the second and third term can be viewed as lower-order diagrams with counter term insertions.

\begin{figure}[t]
\begin{align*}
 &\parbox{3.1cm}{\includegraphics[scale=.3]{1overN2_1}}
 \quad
 \parbox{3.1cm}{\includegraphics[scale=.3]{1overN2_2}}
 \quad
 \parbox{3.1cm}{\includegraphics[scale=.3]{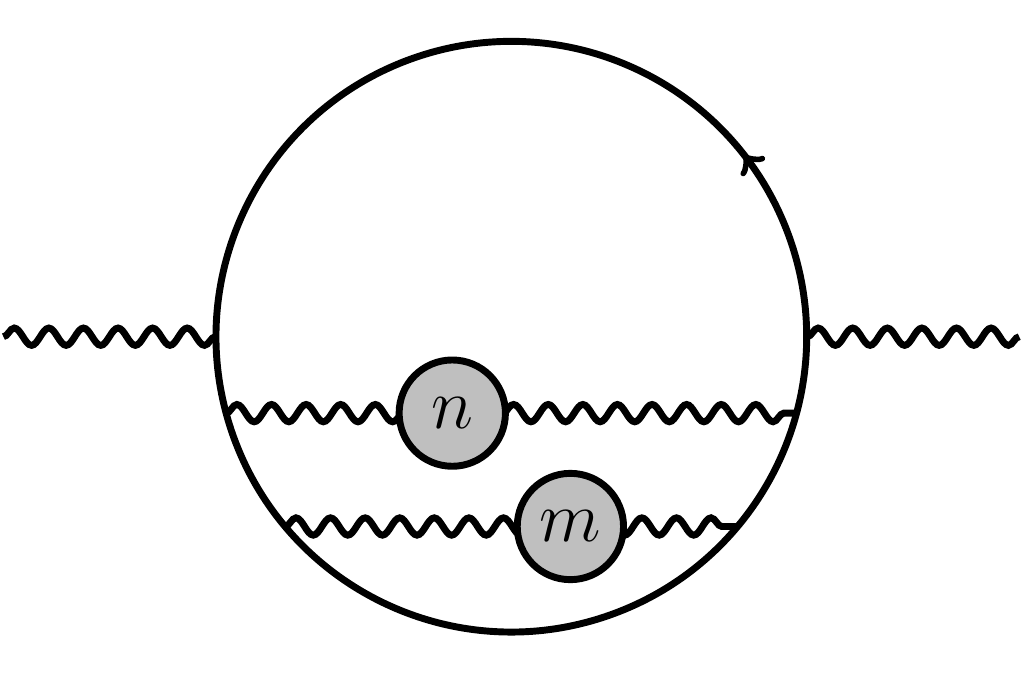}}
 \quad
 \parbox{3.1cm}{\includegraphics[scale=.3]{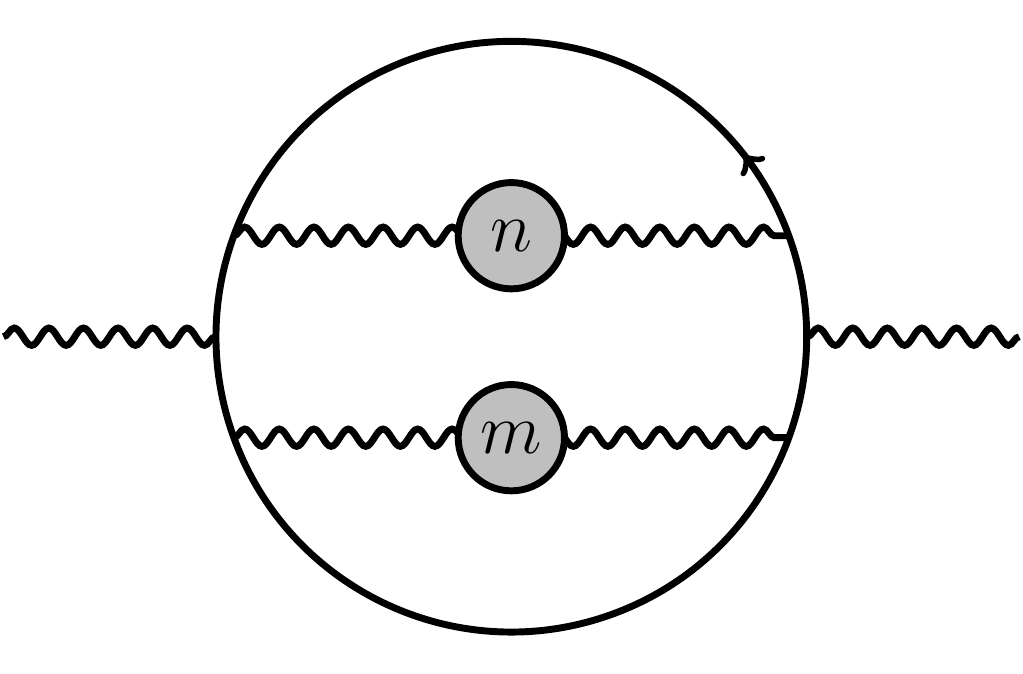}}
  \quad
 \parbox{3.1cm}{\includegraphics[scale=.3]{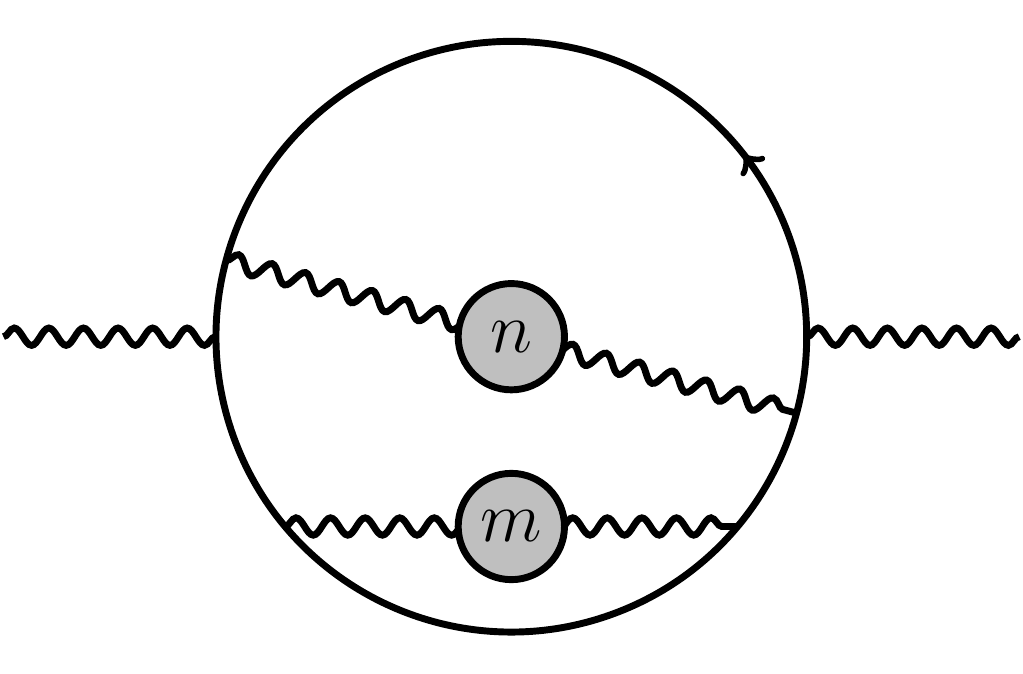}}
 \\
 &\parbox{3.1cm}{\includegraphics[scale=.3]{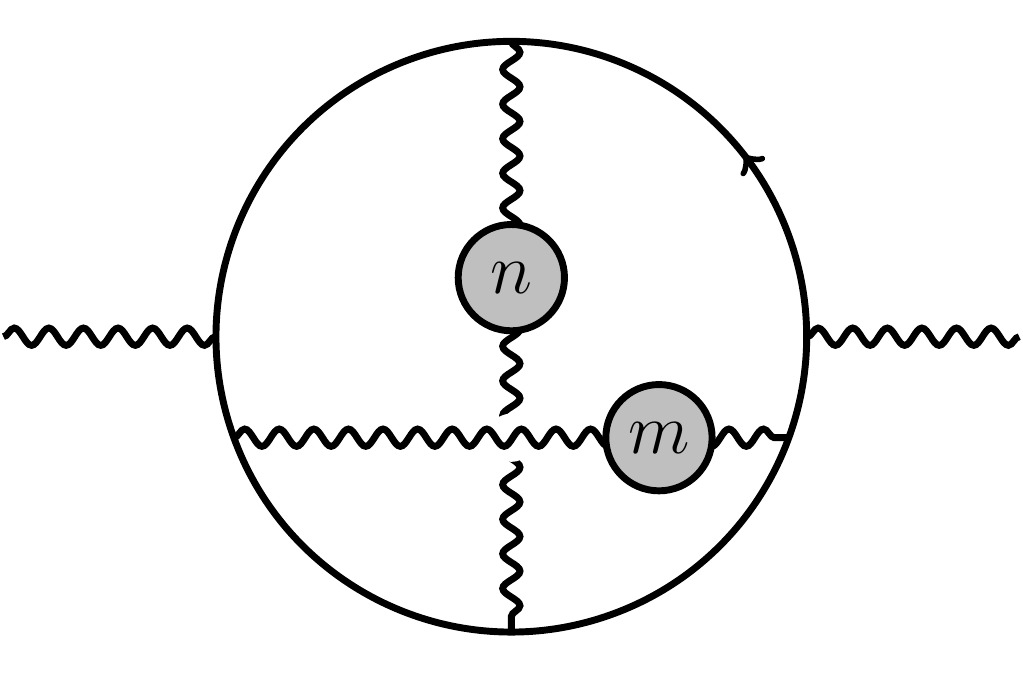}}
 \quad
 \parbox{3.1cm}{\includegraphics[scale=.3]{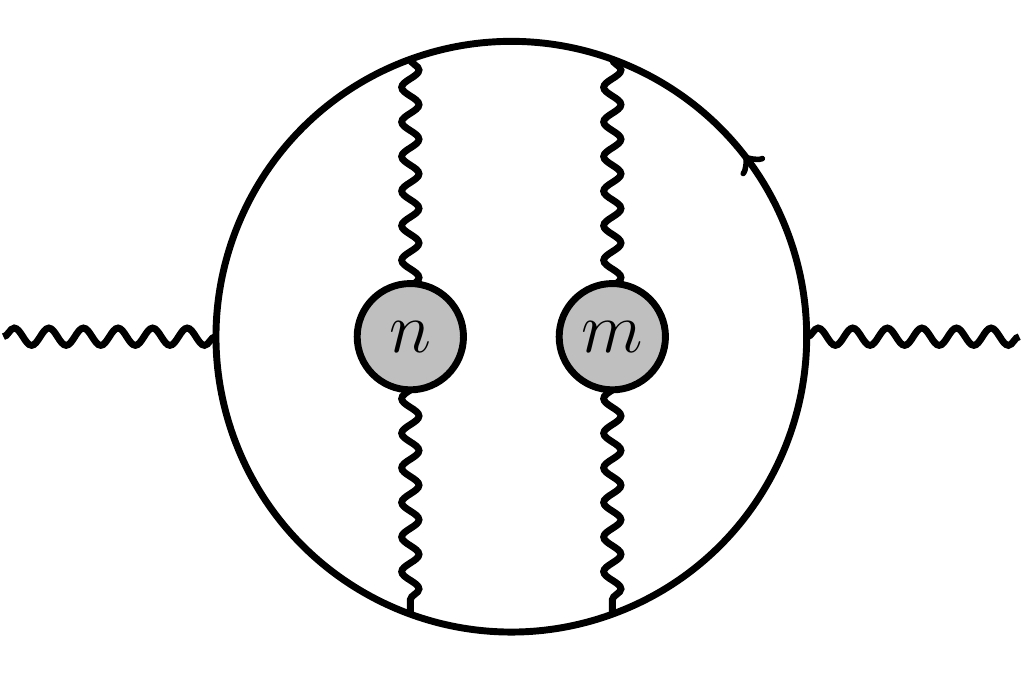}}
 \quad
 \parbox{3.1cm}{\includegraphics[scale=.3]{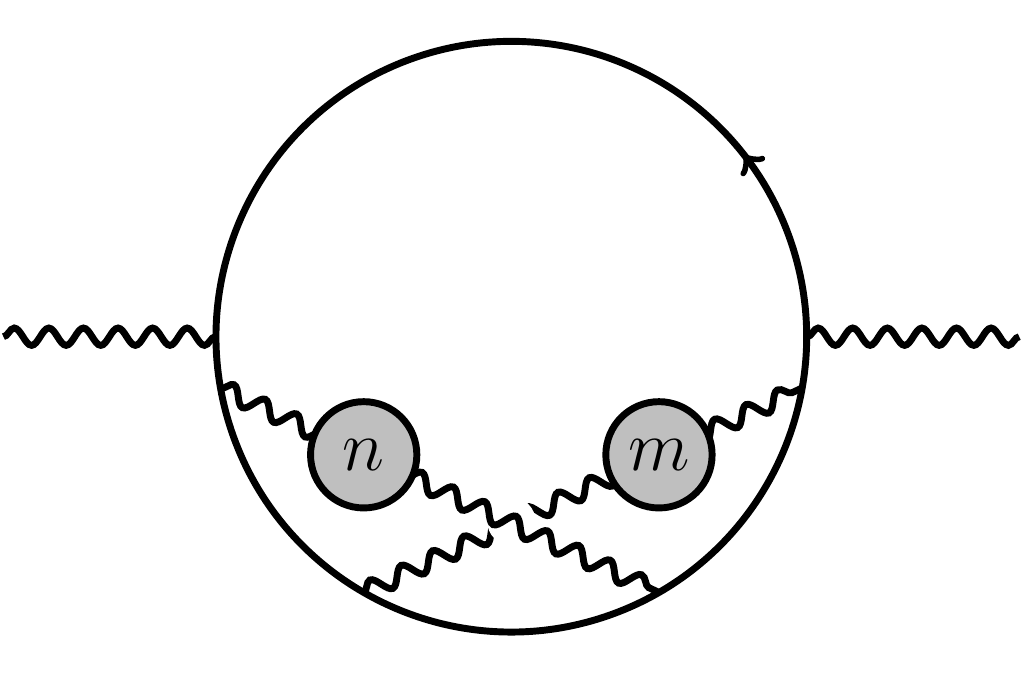}}
 \quad
 \parbox{3.1cm}{\includegraphics[scale=.3]{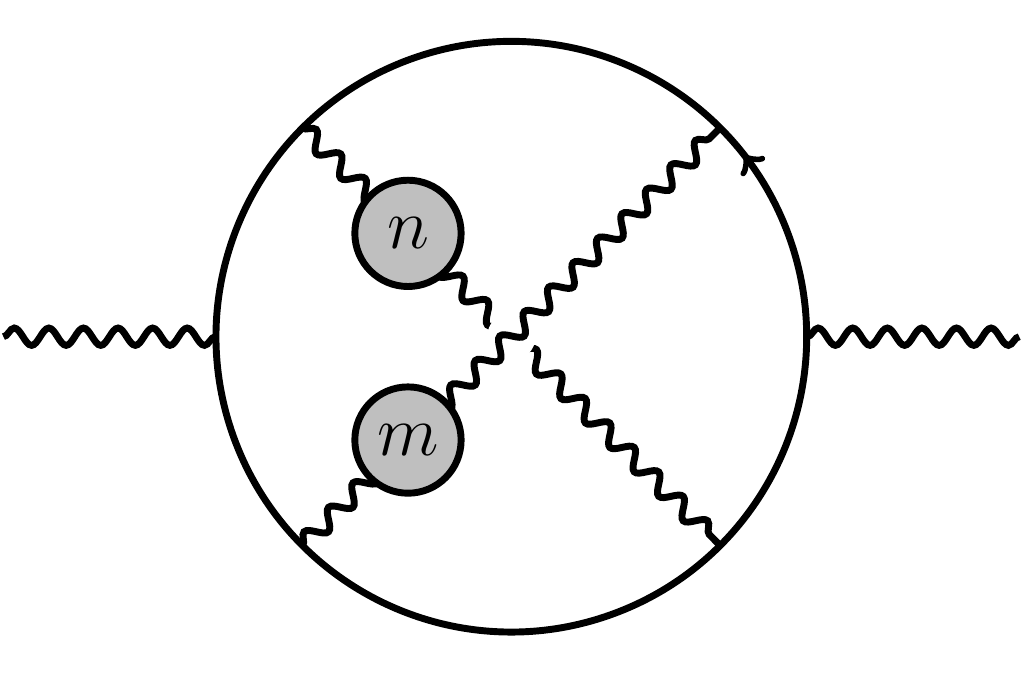}}
 \quad 
 \parbox{4.3cm}{\includegraphics[scale=.27]{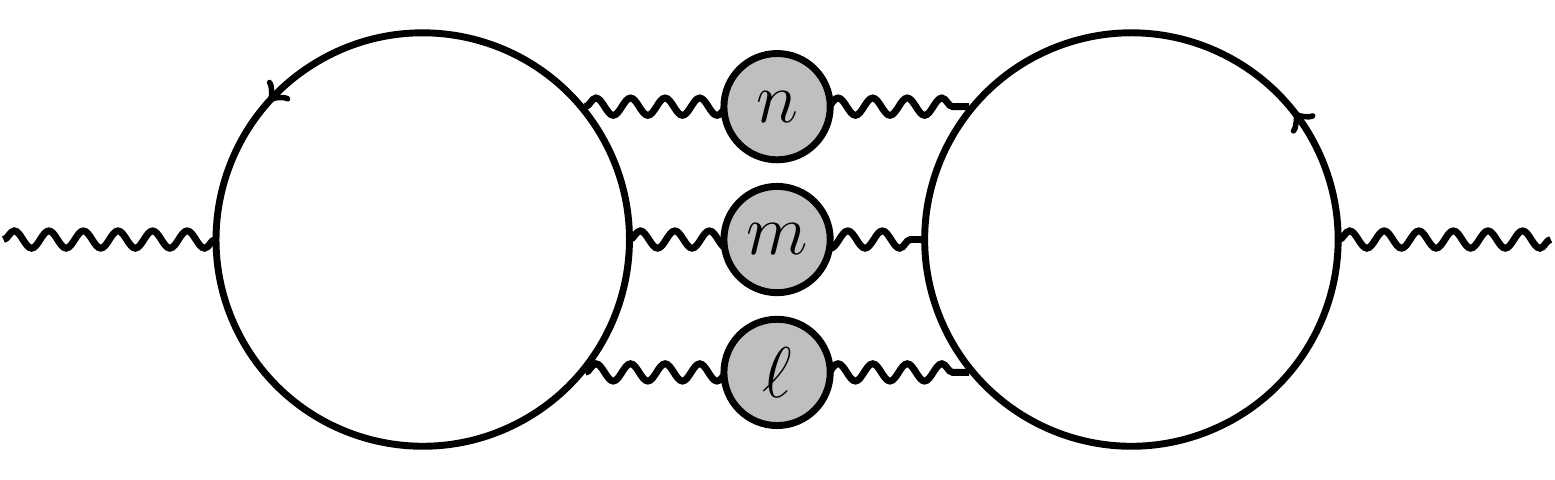}}
 \end{align*}
\caption{Topologies contributing to the beta function at $1/N_f^2$.}
\label{fig:beta2-diags}
\end{figure}

\subsection{Nested diagrams}
We now display the structure of the nested diagrams.
For this it is useful to write the $1/N_f$ contribution to the photon two-point function, i.e., the diagram given in \eqref{eq:beta1-diags}, in an expansion in loop orders
\begin{align}
\label{eq:Pi1}
\Pi_1 =  K_0^2 \sum_{n=0} (- K_0)^n \Pi_1^{(n)}(\epsilon) G_0(\epsilon)^n \left( - \frac{4\pi \mu^2}{p^2} \right)^{(n+2)\epsilon/2} \,.
\end{align}
Here, $\Pi_1^{(n)}$ corresponds to the contribution with $n$ inserted fermion bubbles, $\mu$ is the RG scale, and $G_0$ is the one-bubble amplitude given by
\begin{align} 
 \label{eq:one-bubble}
 G_0 (\epsilon) = 2\frac{\Gamma^2\!\left(2-\frac{\epsilon}{2}\right)\Gamma\!\left(\frac{\epsilon}{2}\right)}{\Gamma\!\left(4-\epsilon\right)} \,.
\end{align}
This notation allows us to write down the nested amplitudes in a convenient way.
For the $1/N_f^2$ nested diagrams, displayed in \eqref{eq:beta2-diags}, we assign $n$ and $m$ fermion bubbles to the outer photon propagators and $\ell$ fermion bubbles to the inner photon propagator.
Then the amplitude is given by
\begin{align}
\label{eq:beta2-amplitude}
\Pi_{2,\text{nested}} &= K_0^4 \sum_{\ell,m,n=0}^{\infty} (-K_0)^{\ell+m+n} \Pi_1^{(\ell)} \Pi_1^{(\ell+m+n+2)} G_0(\epsilon)^{\ell+m+n} \left( - \frac{4\pi \mu^2}{p^2} \right)^{(\ell+m+n+4)\epsilon/2} \notag \\
&= p^2 K_0^4 \sum_{\ell,k=0}^{\infty} (k+1) (-K_0)^{\ell+k} \Pi_1^{(\ell)} \Pi_1^{(\ell+k+2)} G_0(\epsilon)^{\ell+k} \left( - \frac{4\pi \mu^2}{p^2} \right)^{(k+\ell+4)\epsilon/2} \,,
\end{align}
where we used $\sum_{m,n=0} f(m+n) = \sum_{k=0}(k+1)f(k)$.
In straight analogy, we write down the nested amplitudes for $1/N_f^3$.
For the two diagrams in the first line of \eqref{eq:beta3-diags}, the amplitude reads
\begin{align}
\label{eq:beta3-amplitude1}
\Pi_{3,\text{nested},\text{diag1}} &=  K_0^6 \sum_{\ell,m,n,p,q=0}^{\infty}  \Pi_1^{(p)}\Pi_1^{(q)}\Pi_1^{(\ell+m+n+p+q+4)} G_0(\epsilon)^{\ell+m+n+p+q} \left( - \frac{4\pi \mu^2}{p^2} \right)^{(\ell+m+n+p+q+6)\epsilon/2} \notag \\
&=  K_0^6 \sum_{k,p,q=0}^{\infty} \frac{1}{2}(k+1)(k+2) \Pi_1^{(p)}  \Pi_1^{(q)} \Pi_1^{(k+p+q+4)} G_0(\epsilon)^{k+p+q} \left( - \frac{4\pi \mu^2}{p^2} \right)^{(k+p+q+6)\epsilon/2} \,,
\end{align}
where we used $\sum_{\ell,m,n=0} f(\ell+m+n) = \sum_{k=0}\frac{1}{2}(k+1)(k+2) f(k)$.
The two diagrams in the second line of \eqref{eq:beta3-diags} result in the amplitude
\begin{align}
\label{eq:beta3-amplitude2}
\Pi_{3,\text{nested},\text{diag2}} &=  K_0^6 \sum_{l,m,n,p,q=0}^\infty \Pi_1^{(l)} \Pi_1^{(l+m+n+2)} \Pi_1^{(l+m+n+p+q+4)} G_0(\epsilon)^{l+m+n+p+q} \left( - \frac{4\pi \mu^2}{p^2} \right)^{(l+m+n+p+q+6)\epsilon/2} \notag\\
&= K_0^6 \sum_{k,l,r=0}^\infty (k+1)(r+1)\Pi_1^{(l)} \Pi_1^{(k+l+2)} \Pi_1^{(k+l+r+4)} G_0(\epsilon)^{k+l+r} \left( - \frac{4\pi \mu^2}{p^2} \right)^{(k+l+r+6)\epsilon/2} \,.
\end{align}
We close this appendix with a short discussion of the diagrams at $\mathcal{O}(1/N_f^2)$ that are not computed in this paper.
The full set of diagrams contributing to the beta function at $\mathcal{O}(1/N_f^2)$ is displayed in \autoref{fig:beta2-diags}.
For all diagrams in the first line in \autoref{fig:beta2-diags}, the corresponding master integral is known: They all are topologically still two-loop or even one-loop diagrams.
All diagrams in the second line are topologically three-loop diagrams, except the last one, which is a topological four-loop diagram.
The most challenging diagrams are the last two diagrams in the second line in \autoref{fig:beta2-diags}: The first is a non-planar topological three-loop diagram with two bubble chains, while the second one is a topological four-loop diagram with three bubble chains.

\subsection{Finite parts}
\label{app:finite-parts}
We now detail the computation of the finite part of the regularized two-point function $Z_0 \Pi_1$ and its corresponding Borel transform.
We write the amplitude of the two-point function schematically as
\begin{align}
\label{eq:P1b}
\Pi_1 = \frac{3 K_0}{4} \sum_{n=2}^\infty \left( - \frac{2K_0}{3} \right)^{n-1} \frac{1}{n\epsilon^{n-1}} H(n \epsilon ,\epsilon)\,,
\end{align}
where the function $H$ is regular in $n\epsilon$ for constant $\epsilon$ and in $\epsilon$ for constant $n\epsilon$.
Note that this is a different representation of $\Pi_1$ than in \eqref{eq:Pi1}.
In the following, we use the expansion of $H$ in $n\epsilon$ as well as in $\epsilon$, which we denote by
\begin{align}
  H(n \epsilon ,\epsilon) = \sum_{i,j=0}^\infty (n\epsilon)^i \epsilon^j H_{i,j} \,.
\end{align}
We plug this into \eqref{eq:P1b} and also expand $K_0$ in $\epsilon$. We obtain
\begin{align}
\label{eq:Z0Pi1}
Z_0 \Pi_1 &= \frac{3K}{4}\sum_{n=2}^\infty \left( - \frac{2K_0}{3} \right)^{n-1} \frac{1}{n\epsilon^{n-1}} H(n\epsilon,\epsilon)
 = \frac{3K}{4}\sum_{n=2;\,j,k,\ell=0}^\infty  \left( -\frac{2K}{3} \right)^{n+k-1} \binom{n+k-2}{k}\frac{(-1)^{k}n^{j-1}}{\epsilon^{n-\ell+k-j-1}} H_{j,\ell}  \notag \\
&= \frac{3K}{4} \sum_{m=1;\,\ell,j=0}^\infty  \left( - \frac{2K}{3} \right)^m \epsilon^{\ell+j-m} S(j,m) H_{j,\ell}   \,,
\end{align}
where we introduced
\begin{align}
S(j,m) = \sum_{k=0}^{m-1} \binom{m-1}{k} (-1)^k (m-k+1)^{j-1}  \,.
\end{align}
This is computed as 
\begin{align}
S(0,m) &= \frac{(-1)^{m+1}}{m(m+1)} \,,
&
S(m,m) &= (m-1)! \,, 
&
S(j,m) &=0 \quad \forall\, 1 \leq j < m \,.
\end{align}
We denote the finite part, i.e., the limit $\epsilon \rightarrow 0$, of the amplitude \eqref{eq:Z0Pi1} as $\mathcal{F}^{(1)}$. This can be computed as 
\begin{align}
\mathcal{F}^{(1)} \equiv (Z_0 \Pi_1)|_{\epsilon^0} &= \frac{3K}{4} \sum_{m=1}^\infty \sum_{j=0}^m \left(  - \frac{2K}{3} \right)^{m}  S(j,m) H_{j,m-j} \notag \\
&= \frac{3K}{4} \sum_{m=1}^\infty \left(  - \frac{2K}{3} \right)^{m} \left[ \frac{(-1)^{m+1}}{m(m+1)} H_{0,m}  + (m-1)!  H_{m,0} \right] \,.
\end{align}
The first part is a convergent series, while the second one is asymptotic.
For this reason, only the second part can contribute to singularities in the corresponding Borel transform.
We define the Borel transform here by
\begin{align}
\mathcal{F}^{(1)}(K) = \sum_{n=0}^\infty a^{(1)}_n K^{n+2}\qquad \longrightarrow \qquad \mathcal{B}[\mathcal{F}^{(1)}](t) = \sum_{n=0}^\infty \frac{a^{(1)}_n}{n!} t^n \,.
\end{align}
We then obtain 
\begin{align}
\mathcal{B}\!\left[ \mathcal{F}^{(1)}\right] &= \frac{3}{4} \sum_{m=0}^\infty \left(-\frac{2}{3}\right)^{m+1} t^m  H_{m+1,0} + \text{regular} 
\notag \\&
= \frac{3}{4t} \left[ H\!\left(- \frac{2t}{3},0 \right) - H(0,0) \right] +  \text{regular}\,.
\label{eq:borel1}
\end{align}
This is the same formula as in \cite{PalanquesMestre:1985ty} adapted to our notation.
The function $H$ is given by
\begin{align}
H\!\left(- \frac{2t}{3},0 \right)  = \frac{8 e^{\frac{1}{9} (3 \gamma -5) t} M^{-\frac{t}{3}}}{(t+3) (t+6)} &\left(27 \left(\frac{1}{(t-3)^2}-\frac{1}{t^2}+\frac{1}{(t+3)^2}-\frac{1}{(t+6)^2}\right)-\frac{81 \, _3F_2\left(1,2,2-\frac{t}{3};3-\frac{t}{3},3-\frac{t}{3};1\right)}{(t-6)^2 (t-3)}\right. \notag \\
&\quad\left.+3 \pi ^2 \cot \left(\frac{\pi  t}{3}\right) \csc \left(\frac{\pi  t}{3}\right)\right) \,,
\end{align}
with $H(0,0)=1$ and $M=-\frac{4\pi \mu^2}{p^2}$.
Note that the singularities at $t=0$ and $t=3$ in the first and third term cancel.
Only the term with the hypergeometric function is contributing to the singularity at $t=3$.
We display the Borel transform of the finite part in \autoref{fig:Borel-finite-part}.

The finite parts of the $1/N_f^2$ and $1/N_f^3$ contributions are defined as
\begin{align}
	\mathcal{F}^{(2)} ={}& \Bigg\{Z_0 \Pi_2( Z_0^{-1} K) +  Z_1 \left[ 1 - K_0 \frac{\partial}{\partial K_0} \right] \Pi_1(K_0) \Bigg\}_{\epsilon^0} \,, \notag \\
	\mathcal{F}^{(3)}  ={}& \Bigg\{ Z_0 \Pi_3( Z_0^{-1} K)  +  Z_1 \left[ 1 - K_0 \frac{\partial}{\partial K_0} \right] \Pi_2(K_0) + \left( Z_2 - K \frac{Z_2}{Z_0} \frac{\partial}{\partial K_0} + \frac{K^2}{2} \frac{Z_1^2}{Z_0^3} \frac{\partial^2}{\partial K_0^2} \right) \Pi_1(K_0) \Bigg\}_{\epsilon^0} \,.
	\label{eq:finite_parts_def}
\end{align}
No closed-form resummation is possible for these contributions. 
We perform the series expansion in $K$ numerically and provide the coefficients in App.~\ref{app:beta-function}.
We computed the nested part of $\mathcal{F}^{(2)}$ up to $K^{32}$
and the nested part of $\mathcal{F}^{(3)}$ up to $K^{28}$.
The Borel transforms are obtained by dividing out the leading factorial growth.
This leads to the following definition of the respective Borel transforms 
\begin{align}
\mathcal{F}^{(2)}_\text{nested}  &= \sum_{n=0}^{28} a^{(2)}_n K^{n+4}
\qquad \longrightarrow\qquad
\mathcal{B}[\mathcal{F}^{(2)}_\text{nested}](t) = \sum_{n=0}^{28} \frac{a^{(2)}_n}{\Gamma(n+4+1/4)} t^{n} \,, \notag \\
\mathcal{F}^{(3)}_\text{nested}  &= \sum_{n=0}^{22} a^{(3)}_n K^{n+6}
\qquad \longrightarrow\qquad
\mathcal{B}[\mathcal{F}^{(3)}_\text{nested}](t) = \sum_{n=0}^{22} \frac{a^{(3)}_n}{\Gamma(n+6+3/4)} t^n \,.
\end{align}
These Borel transforms are analytically continued using the Conformal-Pad\'e method described in App.~\ref{app:pade}
and plotted in \autoref{fig:Borel-finite-part2} and \autoref{fig:Borel-finite-part3}.

\section{Extracting Physical Quantities from the Large-order Behavior of Perturbative Expansion Coefficients}
\label{app:Darboux}
A fundamental result in complex analysis (Darboux's theorem) states that for a convergent series generated as an expansion at the origin (for example), the large-order behavior of the expansion coefficients is related to the behavior of the function in the vicinity of its singularities. The singularity closest to the origin determines the radius of convergence, and further finer details of the behavior of the function near this singularity are encoded in the sub-leading large-order  behavior of the expansion coefficients at the origin \cite{Fisher,Henrici,gaunt1974asymptotic}. Concretely, 
if a function $f(z)$ has the following branch cut expansion near a singularity $z_0$,
\begin{align}
f(z)\sim  \phi(z)\, \left(1-\frac{z}{z_0}\right)^{-p}+\psi(z) \,, \qquad z\to z_0 \,,
\label{eq:darboux1}
\end{align}
where $\phi(z)$ and $\psi(z)$ are analytic near $z_0$,
then the Taylor expansion coefficients of $f(z)$ at the origin have large-order growth
\begin{align}
b_n \sim{}& \frac{1}{z_0^n}\begin{pmatrix}
n+p-1\\ n
\end{pmatrix} \left[ \phi(z_0)- 
\frac{(p-1)\, z_0\, \phi^\prime(z_0)}{(n+p-1)}  +
\frac{(p-1)(p-2)\, z_0^2\, \phi^{\prime\prime}(z_0)}{2! (n+p-1)(n+p-2)}\, -\dots \right] \,,
\label{eq:darboux2}
\end{align}
If the singularity is logarithmic,
\begin{align}
f(z)\sim  \phi(z)\, \ln\left(1-\frac{z}{z_0}\right)+\psi(z) \,, \qquad z\to z_0 \,,
\label{eq:darboux3}
\end{align}
where $\phi(z)$ and $\psi(z)$ are analytic near $z_0$, 
then the Taylor expansion coefficients of $f(z)$ at the origin have large-order growth
\begin{align}
b_n\sim \frac{1}{z_0^n} \frac{1}{n} \left[\phi(z_0) - \frac{z_0\, \phi^\prime(z_0)}{(n-1)} +\frac{z_0^2\, \phi^{\prime\prime}(z_0)}{(n-1)(n-2)} -\dots \right]\,,
\label{eq:darboux4}
\end{align}
These results can be used in reverse to find the singularity location $z_0$, the exponent $p$ (or to detect logarithmic behavior), and properties of the coefficient function $\phi(z)$, from the large-order growth of the expansion coefficients at the origin. 

\begin{figure}[t]
\includegraphics[width=.49\linewidth]{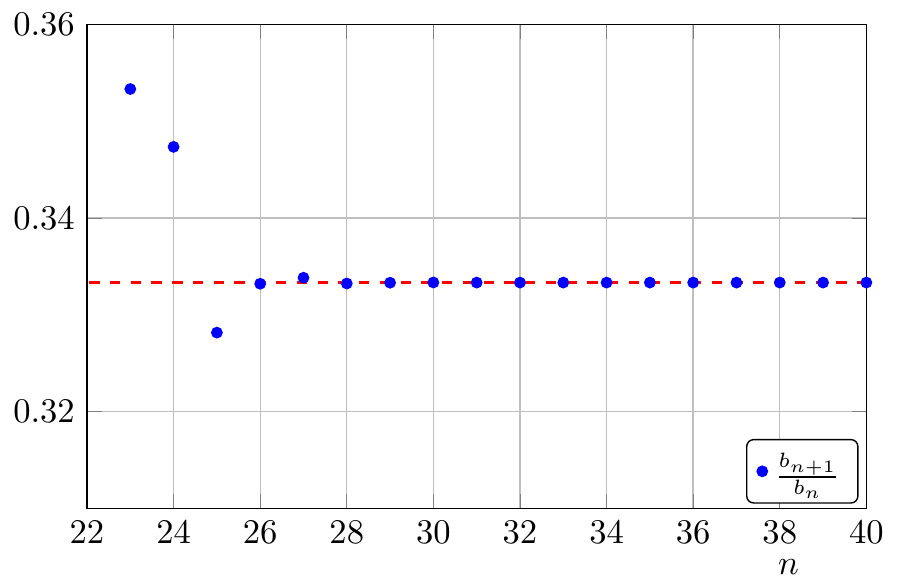} \hfill
\includegraphics[width=.49\linewidth]{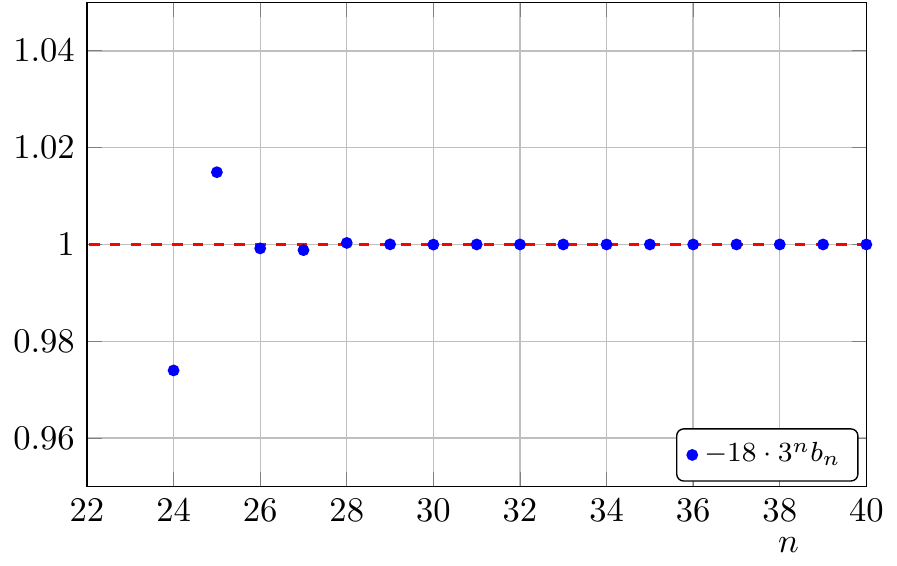} 
\caption{Large-order behavior {\it after taking two derivatives} of the nested beta function.
Left: The ratio test reveals that the radius of convergence is $K=3$.
Right: The prefactor of the leading large-order behavior is determined to be $-\frac{1}{18\cdot3^n}$, indicating a simple pole at $K=3$.
}
\label{fig:derivatives}
\end{figure}

We implemented this strategy on the perturbative expansion of $\beta^{(2)}_\text{nested}(K)$, with expansion coefficients $b_n$. A simple ratio test suggests that $b_{n+1}/b_n\to 1/3$, see \autoref{fig:richardson}. This can be refined using Richardson extrapolation to accelerate the convergence of the ratio test. 
Richardson extrapolation is based on the ansatz \cite{Bender}
\begin{align}
 a_n  = a + \frac{A}{n} + \frac{B}{n^2} + \frac{C}{n^3}  + \dots \,,
\end{align}
where $a$ is the anticipated  convergent value.
First-order Richardson extrapolation is obtained by setting all parameters beyond $1/n$ to zero, i.e., $B=C=\ldots=0$.
The evaluation at $n$ and $n+1$ yields
\begin{align}
 \mathrm R^{(1)} a_n \equiv a = (n+1) a_{n+1} - n a_n \,.
\end{align}
Similarly,  second-order Richardson  extrapolation is obtained by setting the parameters beyond $1/n^2$ to zero, i.e., $C=\ldots=0$, yielding
\begin{align}
 \mathrm R^{(2)} a_n \equiv a= \frac{1}{2}\left((n+2)^2 a_{n+2} -2(n+1)^2 a_{n+1} + n^2 a_n \right) \,.
\end{align}
In \autoref{fig:richardson} we display the effects of the second-order Richardson  extrapolation on the enhancement of the convergence of the ratio test series, clearly indicating convergence to $1/3$, indicating the existence of a singularity at $K_*=3$, and hence a radius of convergence equal to $3$.

Given $z_0$, we can now fit the growth of the coefficients $b_n$ to the branch cut forms in \eqref{eq:darboux2} and \eqref{eq:darboux4}. This can be done by studying the sub-leading behavior of the ratio test. This reveals that the ratio behaves as 
\begin{align}
\frac{b_{n+1}}{b_n}\sim \frac{1}{3}-\frac{2}{3n}+\dots\,, \qquad n\to\infty\,,
\label{eq:bn}
\end{align}
where the precise sub-leading coefficient, $-\frac{2}{3}$, can be extracted using Richardson acceleration once again. This indicates logarithmic behavior, as in \eqref{eq:darboux4}. Now we can probe this further to deduce information about the analytic function $\phi(z)$ multiplying the logarithmic branch cut. The result \eqref{eq:bn} implies that $\phi(3)=0$ and $\phi^\prime(3)=1/6$. 
Analysis of further sub-leading corrections indicate that all higher derivatives of $\phi(z)$ vanish at $z=3$. This leads to the result for the logarithmic branch cut in \eqref{eq:branch-cut}. To confirm this result, we plot in \autoref{fig:richardson} a precise test of the deduced large-order behavior of the $b_n$ coefficients in \eqref{eq:large-order}. The agreement is excellent. Note that with 44 coefficients we can clearly distinguish between $b_n\sim -\frac{1}{2}\frac{1}{3^n}\frac{1}{n(n-1)}$, and the cruder estimate $b_n\sim -\frac{1}{2}\frac{1}{3^n}\frac{1}{n^2}$.
An interesting further consistency test of the logarithmic form of the singularity is to differentiate (twice) the nested beta function $\beta^{(2)}_{\rm nested}(K)$, and then apply the Darboux analysis. The resulting function has a simple pole, which is easy to detect with a ratio test, see \autoref{fig:derivatives} for the convergence of the ratio test in this case.

\section{Pad\'e versus Pad\'e-Conformal}
\label{app:pade}
Pad\'e approximants provide well-known analytic continuations of truncated series expansions and are widely used in physical applications \cite{Bender}. It has further been observed empirically that combining Pad\'e approximants with conformal maps often yields further improved precision \cite{LeGuillou:1979ixc,Caliceti:2007ra,Costin:2019xql}. This improved precision is explained and quantified in \cite{Costin:2020hwg}. The Pad\'e-Conformal analytic continuation procedure for a truncated series in the presence of a branch cut is: (i) first, make a conformal transformation from the cut complex plane to the unit disk; (ii) second, re-expand to the same order inside the conformal disk; (iii) third, make a Pad\'e approximation to the resulting series inside the disk; (iv) finally, map back to the original cut plane with the inverse conformal transformation. This procedure is algorithmically straightforward and is provably exponentially more precise than just Pad\'e if there is a cut \cite{Costin:2019xql,Costin:2020hwg}.

\begin{figure}[t]
\includegraphics[width=.5\linewidth]{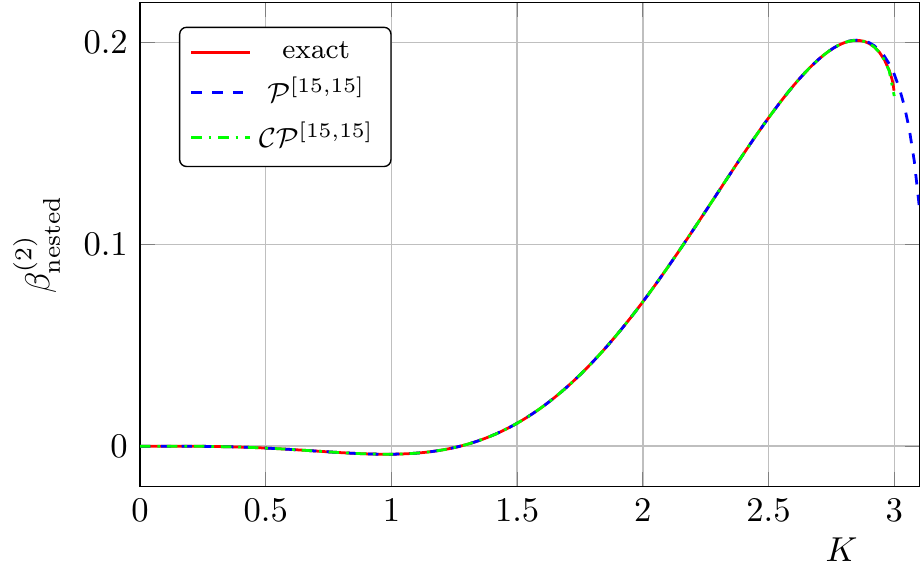}
\caption{Conformal Pad\'e approximant of the nested QED beta function at $\mathcal{O}(1/N_f^2)$ compared to the exact result and a Pad\'e approximant of the same order.}
\label{fig:CPade-1N2}
\end{figure}

In the presence of a single cut (interestingly, it does not matter what the precise {\it nature} of the cut is, just where it is), the explicit conformal map from the $K$ plane cut along the positive real axis with a branch point at $K_*$, together with its inverse are:
\begin{align}
 z = \frac{1-\sqrt{1-\frac{K}{K_*}}}{1+\sqrt{1-\frac{K}{K_*}}}
 \qquad \longleftrightarrow \qquad
 K = \frac{4 K_* z}{(1+z)^2}\,.
\end{align}
The branch cut itself is mapped to the unit circle in the complex $z$ plane.  Given $K_*$, which we have determined to be 3, it is now a completely algorithmic procedure to implement this Pad\'e-Conformal extrapolation. The result is much more precise than just making a Pad\'e approximation, especially in the vicinity of the branch point and branch cut. The results  are shown in  \autoref{fig:pade-1N2} and \autoref{fig:CPade-1N2} for $\beta_\text{nested}^{(2)}(K)$, and in  \autoref{fig:pade-1N3} for $\beta_\text{nested}^{(3)}(K)$.

In the presence of two cuts along the real axis, as occurs for the Borel analysis in \autoref{sec:renormalons},
we use the conformal map from the Borel $t$ plane cut along the positive real axis $t\in [b, \infty)$ and along the negative real axis $t\in (-\infty, -a]$, to the unit disk in the $z$ plane:
\begin{align}
z=\frac{1-\sqrt{\frac{a (b-t)}{b (a+t)}}}{1+\sqrt{\frac{a (b-t)}{b (a+t)}}}
\qquad \longleftrightarrow \qquad
t=\frac{4\, a\, b\, z}{a (1 + z)^2 + b (1 - z)^2} \,.
\end{align}
For the expansions of the finite parts in App.~\ref{app:finite-parts}, the leading singularities are at $t=+3$ and $t=-6$, so we choose $b=3$ and $a=6$.
We map the Borel transform to the unit conformal disk in the $z$ plane, re-expand, and map back again to the Borel $t$ plane.
The resulting plots for the $1/N_f^2$ and $1/N_f^3$ Borel transforms are shown in \autoref{fig:Borel-finite-part2} and \autoref{fig:Borel-finite-part3}, respectively.
Note that the conformal mapping is crucial for revealing the existence of sub-leading Borel singularities~\cite{Costin:2019xql,Costin:2020hwg}.

\section{Beta functions and finite parts}
\label{app:beta-function}
In this appendix, we explicitly display the numerical coefficients of the nested QED beta functions and the finite parts at $\mathcal{O}(1/N_f^2)$ and $\mathcal{O}(1/N_f^3)$. These coefficients are collected in the ancillary Mathematica file with 50 digits of precision. They read
\begin{align}
\beta^{(2)}_\text{nested} ={}& -0.0305 K^4 + 0.0335 K^5 - 0.00335 K^6 - 0.00499 K^7 + 0.00112 K^8 + 
 0.000344 K^9 - 0.000125 K^{10} \notag\\
 & - 9.66\cdot 10^{-6} K^{11} + 7.87\cdot 10^{-6} K^{12} - 
 2.95\cdot 10^{-7} K^{13} - 2.94\cdot 10^{-7} K^{14} + 3.54\cdot 10^{-8} K^{15} + 
 6.29\cdot 10^{-9} K^{16} \notag\\
 &- 1.55\cdot 10^{-9} K^{17} - 3.97\cdot 10^{-11} K^{18} + 
 3.70\cdot 10^{-11} K^{19} - 2.82\cdot 10^{-12} K^{20} - 6.26\cdot 10^{-13} K^{21} \notag\\
 & + 
 4.84\cdot 10^{-14} K^{22} - 9.38\cdot 10^{-15} K^{23} - 4.43\cdot 10^{-15} K^{24} - 
 9.04\cdot 10^{-16} K^{25}- 2.95\cdot 10^{-16} K^{26}  \notag\\
 &- 9.48\cdot 10^{-17} K^{27} - 
 2.89\cdot 10^{-17} K^{28}  - 8.96\cdot 10^{-18} K^{29} - 2.79\cdot 10^{-18} K^{30} - 
 8.70\cdot 10^{-19} K^{31}  \notag\\
 &- 2.72\cdot 10^{-19} K^{32} - 8.52\cdot 10^{-20} K^{33} - 
 2.67\cdot 10^{-20} K^{34} - 8.40\cdot 10^{-21} K^{35} - 2.64\cdot 10^{-21} K^{36}  \notag\\
 &- 
 8.34\cdot 10^{-22} K^{37} - 2.63\cdot 10^{-22} K^{38} - 8.33\cdot 10^{-23} K^{39} - 
 2.64\cdot 10^{-23} K^{40}  - 8.36\cdot 10^{-24} K^{41} \notag\\
 &- 2.65\cdot 10^{-24} K^{42} - 
 8.43\cdot 10^{-25} K^{43} - 2.68\cdot 10^{-25} K^{44} \,,
\end{align}
and
\begin{align}
\beta^{(3)}_\text{nested} ={}&-0.0111 K^6 + 0.0248 K^7 - 0.0113 K^8 - 0.00420 K^9 + 0.00379 K^{10} + 
 0.0000135 K^{11} - 0.000556 K^{12} \notag\\
 &+ 0.0000801 K^{13} + 0.0000461 K^{14} - 
 0.0000128 K^{15} - 2.04\cdot 10^{-6} K^{16} + 1.08\cdot 10^{-6} K^{17} + 
 8.46\cdot 10^{-9} K^{18}\notag\\
 & - 5.85\cdot 10^{-8} K^{19} + 5.18\cdot 10^{-9} K^{20} + 
 2.08\cdot 10^{-9} K^{21} - 3.91\cdot 10^{-10} K^{22} - 4.17\cdot 10^{-11} K^{23} + 
 1.68\cdot 10^{-11} K^{24} \notag\\
 &- 8.47\cdot 10^{-14} K^{25} - 4.73\cdot 10^{-13} K^{26} + 
 4.38\cdot 10^{-14} K^{27} + 9.08\cdot 10^{-15} K^{28} - 1.53\cdot 10^{-15} K^{29} \notag\\
 & +  5.72\cdot 10^{-18} K^{30}+ 5.60\cdot 10^{-17} K^{31} + 3.28\cdot 10^{-18} K^{32} \,.
\end{align}
It is instructive to compare the first few coefficients of these expressions to the complete 5-loop QED $\beta$-function computed in \cite{Baikov:2012zm,Herzog:2017ohr}. This allow us to estimate the nested diagrams contribution to the total result order-by-order in the loop expansion. We find that the $K^4,K^5,K^6$ coefficients of $\beta^{(2)}_\text{nested}$ constitute roughly 50$\%$, 20$\%$, 1$\%$ of the corresponding total 3,4,5-loop coefficient. Moreover, the $K^6$ coefficient of $\beta^{(3)}_\text{nested}$ constitutes less than 1$\%$ of the 5-loop coefficient. This result is expected since the number of diagrams with different topologies that we neglect grows factorially when increasing the loop order. This is also in accord with the fact that the full loop expansion is asymptotic and therefore not convergent.

In \eqref{eq:beta2tilde}, we defined $\tilde \beta^{(2)}_\text{nested} $, which is the nested $\beta$-function at $\mathcal{O}(1/N_f^2)$ with the leading branch cut behavior subtracted.
The coefficients of this function are given by
\begin{align}
\tilde \beta^{(2)}_\text{nested} ={}&
- 0.0300 K^4 + 0.0336 K^5 - 
 0.00333 K^6 - 0.00498 K^7 + 0.00112 K^8 + 0.000344 K^9 - 0.000125 K^{10}  \notag\\
 &- 9.64\cdot10^{-6} K^{11} + 7.88\cdot10^{-6} K^{12} - 
 2.93\cdot10^{-7} K^{13} - 2.93\cdot10^{-7} K^{14}  + 3.55\cdot10^{-8} K^{15} + 
 6.34\cdot10^{-9} K^{16}\notag\\
 & - 1.53\cdot10^{-9} K^{17} - 3.55\cdot10^-{11} K^{18} + 
 3.83\cdot10^{-11} K^{19} - 2.44\cdot10^{-12} K^{20} - 5.13\cdot10^{-13} K^{21}\notag\\
 & + 
 8.29\cdot10^{-14} K^{22}  + 1.11\cdot10^{-15} K^{23} - 1.22\cdot10^{-15} K^{24} + 
 7.98\cdot10^{-17} K^{25} + 7.87\cdot10^{-18} K^{26} \notag\\
 & - 1.39\cdot10^{-18} K^{27} + 
 2.34\cdot10^{-20} K^{28} + 1.05\cdot10^{-20} K^{29} - 8.85\cdot10^{-22} K^{30} - 
 2.15\cdot10^{-23} K^{31} \notag\\
 &+ 7.43\cdot10^{-24} K^{32}  - 3.11\cdot10^{-25} K^{33} - 
 2.62\cdot10^{-26} K^{34} + 3.23\cdot10^{-27} K^{35} - 4.18\cdot10^{-29} K^{36} \notag\\
 &- 
 1.36\cdot10^{-29} K^{37} + 9.06\cdot10^{-31} K^{38}  + 1.30\cdot10^{-32} K^{39} - 
 4.37\cdot10^{-33} K^{40} + 1.61\cdot10^{-34} K^{41} \notag\\
 &+ 8.47\cdot10^{-36} K^{42} - 
 9.53\cdot10^{-37} K^{43} + 1.48\cdot10^{-38} K^{44} \,.
\end{align}
We display the finite parts at the RG scale $\mu = -p^2/(4\pi)$, where $p^2$ is the external momentum.
The finite parts defined as in \eqref{eq:finite_parts_def} at $\mathcal{O}(1/N_f)$ reads
\begin{align}
 \mathcal{F}^{(1)} ={}& 
0.201 K^2 + 0.140 K^3 + 0.0159 K^4 + 0.0726 K^5 + 0.0754 K^6 + 
0.177 K^7 + 0.353 K^8 + 0.957 K^9 + 2.72 K^{10} \notag\\
& + 8.94 K^{11} + 31.9 K^{12} + 126 K^{13} + 536 K^{14} + 2.47\cdot10^3 K^{15} + 1.22\cdot10^4 K^{16} + 6.45\cdot10^4 K^{17} \notag\\
& + 3.62\cdot10^5 K^{18} + 2.15\cdot10^6 K^{19} + 
1.35\cdot10^7 K^{20} + 8.97\cdot10^7 K^{21} + 6.24\cdot10^8 K^{22} + 4.55\cdot10^9 K^{23} \notag\\
& + 3.47\cdot10^{10} K^{24} + 2.76\cdot10^{11} K^{25} + 2.29\cdot10^{12} K^{26} + 
1.98\cdot10^{13} K^{27} + 1.77\cdot10^{14} K^{28} + 1.65\cdot10^{15} K^{29} \notag\\
& + 1.59\cdot10^{16} K^{30} + 1.59\cdot10^{17} K^{31} + 1.63\cdot10^{18} K^{32} + 
1.74\cdot10^{19} K^{33} + 1.91\cdot10^{20} K^{34} + 2.16\cdot10^{21} K^{35} \notag\\
& + 2.51\cdot10^{22} K^{36} + 3.00\cdot10^{23} K^{37} + 3.70\cdot10^{24} K^{38} + 
4.68\cdot10^{25} K^{39} + 6.07\cdot10^{26} K^{40} + 8.08\cdot10^{27} K^{41} \notag\\
& + 1.10\cdot10^{29} K^{42} + 1.54\cdot10^{30} K^{43} + 2.20\cdot10^{31} K^{44} + 3.23\cdot10^{32} K^{45} \,.
\end{align}
The nested contribution at $\mathcal{O}(1/N_f^2)$ reads
\begin{align}
 \mathcal{F}^{(2)}_\text{nested} ={}&
 0.200 K^4 + 0.196 K^5 + 0.369 K^6 + 0.730 K^7 + 1.92 K^8 + 5.11 K^9 + 
 16.2 K^{10} + 55.2 K^{11} + 209 K^{12} + 857 K^{13}\notag\\
 & + 3.80\cdot10^3 K^{14} + 
 1.81\cdot10^4 K^{15} + 9.24\cdot10^4 K^{16} + 5.01\cdot10^5 K^{17 }+ 2.89\cdot10^6 K^{18} + 
 1.76\cdot10^7 K^{19} \notag\\
 &+ 1.13\cdot10^8 K^{20} + 7.65\cdot10^8 K^{21} + 5.43\cdot10^9 K^{22} + 
 4.03\cdot10^{10} K^{23} + 3.12\cdot10^{11} K^{24} + 2.53\cdot10^{12} K^{25} \notag\\
 &+ 
 2.13\cdot10^{13} K^{26} + 1.86\cdot10^{14} K^{27} + 1.69\cdot10^{15} K^{28} + 
 1.60\cdot10^{16} K^{29} + 1.56\cdot10^{17} K^{30} + 1.57\cdot10^{18} K^{31} \notag\\
 &+ 1.64\cdot10^{19} K^{32} \,,
\end{align}
while the nested contribution to the finite part at $\mathcal{O}(1/N_f^3)$ is
\begin{align}
 \mathcal{F}^{(3)}_\text{nested} ={}& 0.300 K^6 + 0.760 K^7 + 2.29 K^8 + 7.71 K^9 + 26.9 K^{10} + 103 K^{11} + 
 423 K^{12} + 1.87\cdot10^3 K^{13} + 8.83\cdot10^3 K^{14} \notag\\
 &+ 4.45\cdot10^4 K^{15} + 
 2.38\cdot10^5 K^{16}+ 1.35\cdot10^6 K^{17} + 8.10\cdot10^6 K^{18} + 5.12\cdot10^7 K^{19} + 
 3.40\cdot10^8 K^{20} \notag\\
 &+ 2.37\cdot10^9 K^{21} + 1.73\cdot10^{10} K^{22} + 
 1.32\cdot10^{11} K^{23} + 1.05\cdot10^{12} K^{24} + 8.67\cdot10^{12} K^{25} + 
 7.46\cdot10^{13} K^{26} \notag\\
 &+ 6.66\cdot10^{14} K^{27}+6.17\cdot 10^{15} K^{28} \,.
\end{align}
The coefficients of all finite parts grow factorially, as expected. From the above expressions for the finite parts we can already see that the factorial rate of growth of the coefficients is comparable at $\mathcal{O}(1/N_f)$, $\mathcal{O}(1/N_f^2)$, and $\mathcal{O}(1/N_f^3)$. Moreover, since all the coefficients are positive, we deduce that the leading Borel singularity should be on the positive real axis. Indeed, our further analysis shows that this leading singularity is at $t=3$, for each order of the large $N_f$ expansion. See \autoref{fig:Borel-finite-part}, \autoref{fig:Borel-finite-part2} and \autoref{fig:Borel-finite-part3}.

\end{widetext}

\bibliography{largeN}

\end{document}